\documentclass[10pt,pra,showpacs,superscriptaddress,nofootinbib,amssymb,amsmath,twocolumn,eqsecnum]{revtex4}
\usepackage[dvips]{graphicx}
\usepackage{color,latexsym,epsfig,bm,times,subfigure}
\usepackage[ps2pdf,bookmarks=true,colorlinks,linkcolor=blue,urlcolor=green,citecolor=red]{hyperref}

\renewcommand\a{\alpha}
\renewcommand\b{\beta}
\renewcommand\d{\delta}

\renewcommand\l{\lambda}

\renewcommand\t{\tau}

\renewcommand\c{\chi}
\renewcommand\j{\psi}
\renewcommand\o{\omega}

\newcommand\g{\gamma}
\newcommand\z{\zeta}
\newcommand\m{\mu}
\newcommand\n{\nu}
\newcommand\x{\xi}
\newcommand\p{\pi}

\newcommand\s{\sigma}

\newcommand\w{\eta}
\newcommand\ve{\varepsilon}

\renewcommand\P{\Pi}
\renewcommand\S{\Sigma}
\renewcommand\O{\Omega}
\renewcommand\H{\Theta}
\newcommand\D{\Delta}
\newcommand\G{\Gamma}

\newcommand\J{\Psi}

\newcommand{\fig}[1]{Fig.~\ref{#1}}
\newcommand{\eq}[1]{Eq.~(\ref{#1})}
\newcommand{\eqs}[2]{Eqs.~(\ref{#1})-(\ref{#2})}

\newcommand\lb{\left(}
\newcommand\rb{\right)}
\newcommand\ls{\left[}
\newcommand\rs{\right]}

\newcommand{\lan}{\langle}
\newcommand{\ran}{\rangle}
\newcommand\ua{\uparrow}
\newcommand\da{\downarrow}
\newcommand\ra{\rightarrow}

\newcommand{\non}{\nonumber\\}
\newcommand\pt{\partial}

\newcommand{\ie}{\emph{i.e.}}

\newcommand{\ch}{{\cal H}}
\newcommand{\cs}{{\cal S}}
\newcommand{\cf}{{\cal F}}
\newcommand{\cg}{{\cal G}}
\newcommand{\ct}{{\cal T}}

\newcommand{\mf}{{\rm mf}}
\renewcommand{\sf}{{\rm sf}}
\newcommand{\pg}{{\rm pg}}
\newcommand{\Tr}{{\rm Tr}}

\newcommand{\im}{{\rm{Im}}}

\newcommand{\br}{{\mathbf r}}

\newcommand{\bk}{{\mathbf k}}
\newcommand{\bq}{{\mathbf q}}

\newcommand{\bcs}{{\rm{BCS}}}
\newcommand{\bec}{{\rm{BEC}}}

\newcommand{\kf}{k_{\rm F}}
\newcommand{\vf}{v_{\rm F}}
\newcommand{\ca}{{\cal A}}
\newcommand{\cb}{{\cal B}}

\begin{document}

\title{BCS-BEC crossover at finite temperature in spin-orbit coupled Fermi gases}
\author{Lianyi He}\email{lianyi@fias.uni-frankfurt.de}
\affiliation{Frankfurt Institute for Advanced Studies and Institute for Theoretical Physics,
J. W. Goethe University, 60438 Frankfurt am Main, Germany}
\author{Xu-Guang Huang}\email{xuhuang@indiana.edu}
\affiliation{Frankfurt Institute for Advanced Studies and Institute for Theoretical Physics,
J. W. Goethe University, 60438 Frankfurt am Main, Germany}
\affiliation{Physics Department and Center for Exploration of Energy and Matter,
Indiana University, 2401 N Milo B. Sampson Lane, Bloomington, IN 47408, USA}
\author{Hui Hu}
\affiliation{ARC Centre of Excellence for Quantum-Atom Optics, Centre for Atom Optics and Ultrafast Spectroscopy,
Swinburne University of Technology, Melbourne 3122, Australia}
\author{Xia-Ji Liu}
\affiliation{ARC Centre of Excellence for Quantum-Atom Optics, Centre for Atom Optics and Ultrafast Spectroscopy,
Swinburne University of Technology, Melbourne 3122, Australia}
\date{\today}

\begin{abstract}
By adopting a $T$-matrix-based method within the $G_0G$ approximation for the pair susceptibility, we
study the effects of the pairing fluctuation on the three-dimensional spin-orbit-coupled Fermi gases at finite temperature.
The critical temperatures of the superfluid to normal phase transition are determined for
three different types of spin-orbit coupling (SOC): (1) the extreme oblate (EO)
or Rashba SOC, (2) the extreme prolate or equal Rashba-Dresselhaus SOC, and (3) the spherical (S) SOC.
For EO- and S-type SOC, the SOC dependence of the critical temperature signals
a crossover from BCS to BEC state; at strong SOC limit, the critical temperature recovers those of ideal BEC of rashbons.
The pairing fluctuation induces a pseudogap in the fermionic excitation spectrum in both superfluid and
normal phases. We find that, for EO- and S-type SOC, even at weak coupling, sufficiently strong SOC can induce
sizable pseudogap.
Our research suggests that the spin-orbit-coupled Fermi gases may
open new means to the study of the pseudogap formation in fermionic systems.
\end{abstract}
\pacs{03.75.Ss, 05.30.Fk, 67.85.Lm, 74.20.Fg}

\maketitle

\section {Introduction}\label{intro}
The experimental realization of ultracold Fermi gases with tunable interatomic interaction has
opened new era for the study of some longstanding theoretical proposals in many-fermion systems.
One particular example is the smooth crossover
from a Bardeen-Cooper-Schrieffer (BCS) superfluid ground state with largely overlapping Cooper
pairs to a Bose-Einstein condensate (BEC) of tightly bound bosonic molecules $-$
a phenomenon suggested many years ago~\cite{Eagles:1969zz,Leggett:1980,Nozieres:1985zz,SadeMelo:1993zz}.
For a dilute Fermi gas in three dimensions (3D) with a short-range
interatomic interaction where the effective
range $r_0$ of the interaction is much smaller than the interatomic distance, such a
BCS-BEC crossover can be characterized by the dimensionless gas parameter, $1/(\kf a)$, where
$\kf$ is the Fermi momentum and $a$ is the $s$-wave scattering length of the short-range interaction.
The BCS-BEC crossover occurs when $1/(\kf a)$ is tuned from negative to positive values (the turning point
is called unitarity).

This BCS-BEC crossover has been successfully demonstrated in ultracold Fermi gases where the $s$-wave
scattering length is tuned by means of the Feshbach resonance~\cite{Greiner:2003zz,Jochim:2003,Zwierlein:2003}.
This has been regarded as one of the key successes
in the cold-atom researches and has attracted broad interests due to its special properties. For example, near unitarity, the system is a high-$T_c$ superfluid: the superfluid to normal transition temperature $T_c$ is much higher than that of an ordinary BCS superfluid. The normal state near unitarity is strongly affected by many-body effects, e.g., the pair fluctuations which we will thoroughly study, leading to deviations from a Fermi liquid behavior and pseudogap opening, as in (underdoped) cuprate superconductors. Curiously, it is always interesting to look for other mechanisms of realizing
the BCS-BEC crossover. Recent experimental breakthrough in generating synthetic non-Abelian gauge field in
bosonic gas of $^{87}$Rb atoms
has opened the opportunity to study the spin-orbit-coupling (SOC) effects in cold atomic gases~\cite{Spielman:2011}. In this experiment, two counter propagating Raman laser beams and a transverse Zeeman field are applied to
$^{87}$Rb atoms, and three hyperfine levels of $^{87}$Rb are coupled by the Raman lasers. By tuning the Zeeman
field energy and the Raman laser frequency, two of the three hyperfine levels can become degenerate (which can be interpreted as two spins) and the low energy physics can be described by a model Hamiltonian of a spinor Bose gas coupled to an external spin $SU(2)$ non-Abelian gauge field which, in their special setup, turns out to induce a SOC. For the
fermionic case, some of the recent theoretical results suggested that tuning the SOC may provide
an alternative way to realize the BCS-BEC
crossover~\cite{Vyasanakere:2011,Hu:2011,Yu:2011,Iskin:2011a,Gong:2011,Han:2011,Yi:2011,He:2011jva,He:2011,other}. The experimental exploration of the spin-orbit coupled Fermi gases has also achieved remarkable progresses and
the Raman scheme designed for generating SOC in $^{87}$Rb atoms has been successfully applied to Fermi gases:
the spin-orbit coupled $^{40}$K
and $^6$Li atoms have been realized at Shanxi University~\cite{Wang:2012} and at Massachusetts Institute of Technology (MIT)~\cite{Cheuk:2012},
respectively.

The SOC of fermions can be induced by a synthetic uniform $SU(2)$ gauge field, $A^\m_i=\l_i\d^\m_i$,
where $\l_i$ will play the roles of the SOC strengths. With this gauge field, the single-particle
Hamiltonian reads $\ch=\bk^2/(2m)-{\bm \s}\cdot{\bf s}(\bk)$ where ${\bf s}(\bk)=(\l_xk_x,\l_yk_y,\l_zk_z)$. In a
very interesting paper~\cite{Vyasanakere:2011b}, Vyasanakere and Shenoy studied the two-body problem of this Hamiltonian. They paid
particular attention to three special types of gauge field configurations: (1) $\l_x=\l_y=0$ and $\l_z=\l$ [called extreme prolate (EP)], (2) $\l_x=\l_y=\l$ and $\l_z=0$ [called extreme oblate (EO)], and (3) $\l_x=\l_y=\l_z$ [
called spherical (S)]. The EO SOC is physically equivalent to the Rashba SOC which has been famous in
condensed matter physics. The EP SOC is physically equivalent to an equal mixture of Rashba and Dresselhaus SOCs.
The most surprising finding of Vyasanakere and Shenoy was that for EO and S SOCs, even
for $a<0$ where the di-fermion bound state cannot form in the absence of SOC, the di-fermion bound state (referred to as rashbon) always
exists and its binding energy is generally enhanced with increased SOC. Meanwhile, the bound state also possesses
non-trivial effective mass which is generally larger than twice of the fermion mass $m$. For the two dimensional (2D) case, although a bound state exists for arbitrarily small attraction, it was shown in Ref.~\cite{He:2011jva} that the EO or Rashba SOC can generally enhance the binding energy and the effective mass of the bound state.

The novel bound state that emerged in the two-body problem suggests that the EO or S SOCs
may trigger a new type of BCS-BEC crossover in the many-body problem of fermions. In fact, theoretical studies revealed that
for EO or S SOC, even at small negative $\kf a$, a crossover from the BCS superfluid to the BEC of rashbons
can be achieved by tuning the SOC $\l$ to large enough value{~\cite{Vyasanakere:2011,Hu:2011,Yu:2011,Iskin:2011a,Gong:2011,Han:2011,Yi:2011,He:2011jva,He:2011,other}. It was shown that for EO or S SOCs the system enters the rashbon
BEC regime at $\l\sim \vf$
where $\vf\equiv \kf/m$ is the Fermi velocity. Similar conclusions were also found for 2D Fermi gases with EO SOC~\cite{He:2011jva}.

So far, most of the theoretical studies of the BCS-BEC crossover in 3D SOC Fermi gases focused on the
zero-temperature ground state based on mean-field theory (MFT). Although the MFT captures some qualitative features
of the zero-temperature crossover, it loses the effects of the pairing fluctuation which becomes substantial
when the system goes toward finite-temperature and/or the BEC regimes.
In the absence of the SOC, previous theoretical studies~\cite{Loktev:2000ju,PhysRevB58.R5936,phyc1999,prb61:11662,phyrept2005,arXiv:0810.1938,Chien:2009,PG1,strinati} as well as quantum Monte Carlo simulation~\cite{PG3} have already revealed that,
as a consequence of the pairing fluctuation, a ``pseudogap" emerges in the fermionic excitation
spectrum.
This
pseudogap is negligibly small at BCS limit but increases as $1/(\kf a)$ is increased and becomes significantly
important on the BEC side. Particularly, the pseudogap survives above the superfluid critical temperature $T_c$ and
leads to an exotic normal state that is different from the Fermi-liquid normal state associated with the MFT.
Recently, the experimental observation of pairing pseudogap in both 2D Fermi gases~\cite{Feld:2011} and 3D Fermi gases~\cite{PG2} are reported. Similar
pseudogap phenomena may also appear in other strongly correlated systems, such as high-$T_c$ superconductors~\cite{Loktev:2000ju,Chen:1998zz,htc,Damascelli},
low-density nuclear matter~\cite{Huang:2010}, and color superconducting quark matter~\cite{csc}.

In this paper, we study the spin-orbit-coupled Fermi gases at finite temperature. To include the pairing-fluctuation
effects and investigate the possible pseudogap phenomena, we will adopt a $T$-matrix formalism based on a
$G_0G$ approximation for the pair susceptibility which was first introduced by the Chicago group~\cite{PhysRevB58.R5936,phyc1999,prb61:11662,phyrept2005,arXiv:0810.1938,Chien:2009,Chen:1998zz}. This formalism
generalizes the early works of Kadanoff and Martin~\cite{km} and Patton~\cite{patton}, and can be considered as a natural extension of
the BCS theory since they share the same ground state. Moreover, this formalism allows quasi-analytic
calculations and gives a simple physical interpretation of the pseudogap emergence. It clearly shows that the
pseudogap is due to the incoherent pairing fluctuation. Within this formalism, we can also determine
the superfluid critical temperature and study how the pairing fluctuation affects the thermodynamics. We note that
this is the first systematic study of the 3D spin-orbit-coupled Fermi gases at finite temperature. For 2D spin-orbit-coupled Fermi gases, the possible BKT transition at finite temperature was already studied~\cite{He:2011jva}.

The article is organized as follows. In Sec. \ref{tmat}, we present a detailed theoretical scheme of the $T$-matrix-based formalism at
finite temperature. The numerical results are given in Sec. \ref{resul}. We summarize in Sec. \ref{discu}. Throughout this article, we use natural units $\hbar=k_B=1$.

\section {T-matrix-based formalism}\label{tmat}
We consider a homogenous Fermi gas interacting via a short-range attractive interaction in the spin-singlet channel. In the
dilute limit $\kf r_0\ll1$ and $m\l r_0\ll 1$, where $r_0$ is the effective interaction range\footnote{The interaction range $r_0$ is about 3.2 nm for $^{40}$K~\cite{K40} and 2.1 nm for $^6$Li~\cite{Li6}. Thus for these atoms when $\kf, m\l\gtrsim0.1$ nm$^{-1}$ the dilute condition may be violated. In Shanxi University experiment~\cite{Wang:2012}, $m\l=0.008$ nm$^{-1}$ and $\kf$ varies from $0.9m\l$ to $1.8m\l$; in MIT experiment~\cite{Cheuk:2012}, $m\l=0.003$ nm$^{-1}$ and $\kf$ varies about $m\l$. In both experiments, the dilute conditions are well satisfied.}, this system can be
described by the following Hamiltonian,
\begin{eqnarray}
\label{lagr}
H &=&\int d^3 {\bf r}\psi^{\dagger}({\bf r}) \left({\cal
H}_0+\cal{H}_{\rm{so}}\right) \psi({\bf r})\nonumber\\
&+&U\int d^3 {\bf r}^{\phantom{\dag}}\psi^\dagger_{\uparrow}({\bf
r})\psi^\dagger_{\downarrow}({\bf
r})\psi^{\phantom{\dag}}_{\downarrow}({\bf
r})\psi^{\phantom{\dag}}_{\uparrow}({\bf r}),
\end{eqnarray}
where $\ch_0=-\nabla^2/(2m)-\m$ is the free single-particle Hamiltonian with $\m$ being the chemical potential,
$\ch_{\rm so}=-i\sum_{i=1}^3\l_i\s_i\pt_i$ is the SOC term, and $U<0$ denotes the attractive $s$-wave interaction.

Introduce the four-dimensional Nambu-Gorkov spinor $\J=(\j_\ua, \j_\da, \j_\ua^\dag, \j_\da^\dag)^T$.
The (imaginary-time)
Green's function of the Nambu-Gorkov spinor is given by
\begin{eqnarray}
\cs(\t,\br)&\equiv&-\lan T_\t\J(\t,\br)\J^\dag(0,{\bf 0})\ran\non
&=&\ls\begin{matrix}
\cg(\t,\br) & \cf(\t,\br) \\ \tilde{\cf}(\t,\br) & \tilde{\cg}(\t,\br)
\end{matrix}\rs,
\end{eqnarray}
where $T_\t$ is the (imaginary-) time-ordering operator and $\t\equiv it$.
It is convenient to work in frequency-momentum space,
\begin{eqnarray}
\cs(K)&=&\ls\begin{matrix}
\cg(K) & \cf(K) \\ \tilde{\cf}(K) & \tilde{\cg}(K)
\end{matrix}\rs,
\end{eqnarray}
where $K=(k_0\equiv i\o_n, \bk)$ with $\o_n=(2n+1)\p T$ ($n$ integer) being the Matsubara frequency for fermion.
The Green's functions have the following properties:
\begin{eqnarray}
\label{g1}
\tilde{\cg}(i\o_n,\bk)&=&-[\cg(-i\o_n,-\bk)]^T,\\
\label{g2}
\tilde{\cf}(i\o_n,\bk)&=&+[\cf(-i\o_n,\bk)]^\dag,\\
\label{g3}
\cf(i\o_n,\bk)&=&-[\cf(-i\o_n,-\bk)]^T,\\
\label{g4}
\tilde{\cf}(i\o_n,\bk)&=&-[\tilde{\cf}(-i\o_n,-\bk)]^T,\\
\label{g5}
\cg(i\o_n,\bk)&=&+[\cg(-i\o_n,\bk)]^\dag,\\
\label{g6}
\tilde{\cg}(i\o_n,\bk)&=&+[\tilde{\cg}(-i\o_n,\bk)]^\dag.
\end{eqnarray}

In the rest of this section, we will introduce the basic method of the $T$-matrix. Our strategy will closely follow Refs.~\cite{PhysRevB58.R5936,phyc1999,prb61:11662,phyrept2005,arXiv:0810.1938,Chien:2009,Chen:1998zz,Huang:2010}. The $T$-matrix we will adopt is defined as an
infinite series of ladder-diagrams in the particle-particle channel by constructing the ladder by one free particle propagator
and one full particle propagator. The $T$-matrix thus enters the particle self-energy in place of the bare interaction vertex.
The equation that defines the $T$-matrix, the self-energy equation (or gap equation) as well as the number density equation form
a closed set of equations, and should be solved consistently. One can view this approach as the simplest generalization of
the BCS theory, which can also be cast formally into a $T$-matrix formalism.
Let us start with the BCS theory.

\subsection {BCS Theory}\label{bcs}
The BCS theory is based on the mean-field approximation to the anomalous self-energy. We start with the mean-field inverse fermion
propagator,
\begin{eqnarray}
\cs^{-1}_\mf(K)
&=&\ls\begin{matrix}
\cg_0^{-1}(K) & i\s_y\D_\mf \\ -i\s_y\D_\mf & -[\cg_0^{-1}(-K)]^T
\end{matrix}\rs,
\end{eqnarray}
where the anomalous self-energy $\D_\mf$ is chosen as a constant and can be used as an order parameter for superfluid phase transition.
The inverse free fermion propagator reads,
\begin{eqnarray}
\cg_0^{-1}(K)=i\o_n-\x_\bk-\x_{\rm so}(\bk),
\end{eqnarray}
with $\x_\bk=\bk^2/(2m)-\m$ and $\x_{\rm so}(\bk)=\sum_{i=1}^3\l_i\s_ik_i$ ($\l_i$ is real). By direct doing the matrix inversion, one obtains
\begin{eqnarray}
\label{mfs}
\cs_\mf(K)&=&\ls\begin{matrix}
\cg_\mf(K) & \cf_\mf(K) \\ \tilde{\cf}_\mf(K) & \tilde{\cg}_\mf(K)
\end{matrix}\rs.
\end{eqnarray}
Its elements are
\begin{eqnarray}
\label{mfs1}
\cg_\mf(K)&=& \ca_{11}(K)+\frac{\xi_{\rm so}(\bk)}{\l|\bk|}\cb_{11}(K),\\
\label{mfs2}
\tilde{\cg}_\mf(K)&=& \ca_{22}(K)+\frac{\xi_{\rm so}^*(\bk)}{\l|\bk|}\cb_{22}(K),\\
\label{mfs3}
\cf_\mf(K)&=&-i\s_y\ls\ca_{12}(K)+\frac{\xi_{\rm so}^*(\bk)}{\l|\bk|}\cb_{12}(K)\rs,\\
\label{mfs4}
\tilde{\cf}_\mf(K)&=&i\s_y\ls\ca_{21}(K)+\frac{\xi_{\rm so}(\bk)}{\l|\bk|}\cb_{21}(K)\rs,
\end{eqnarray}
where we introduced $\l|\bk|\equiv\sqrt{\sum_{i=1}^3\l_i^2k_i^2}$ and
\begin{eqnarray}
&&{\cal A}_{11}(K)=\frac{1}{2}\left[\frac{i\omega_n+\xi_{\bf
k}^+}{(i\omega_n)^2-(E_{\bf k}^+)^2}+\frac{i\omega_n+\xi_{\bf
k}^-}{(i\omega_n)^2-(E_{\bf k}^-)^2}\right],\nonumber\\
&&{\cal A}_{22}(K)=\frac{1}{2}\left[\frac{i\omega_n-\xi_{\bf
k}^+}{(i\omega_n)^2-(E_{\bf k}^+)^2}+\frac{i\omega_n-\xi_{\bf
k}^-}{(i\omega_n)^2-(E_{\bf k}^-)^2}\right],\nonumber\\
&&{\cal A}_{12}(K)=\frac{1}{2}\left[\frac{\Delta_\mf}{(i\omega_n)^2-(E_{\bf
k}^+)^2}+\frac{\Delta_\mf}{(i\omega_n)^2-(E_{\bf
k}^-)^2}\right],\nonumber\\
&&{\cal A}_{21}(K)={\cal A}_{12}(K)
\end{eqnarray}
and
\begin{eqnarray}
&&{\cal B}_{11}(K)=\frac{1}{2}\left[\frac{i\omega_n+\xi_{\bf
k}^+}{(i\omega_n)^2-(E_{\bf k}^+)^2}-\frac{i\omega_n+\xi_{\bf
k}^-}{(i\omega_n)^2-(E_{\bf k}^-)^2}\right],\nonumber\\
&&{\cal B}_{22}(K)=-\frac{1}{2}\left[\frac{i\omega_n-\xi_{\bf
k}^+}{(i\omega_n)^2-(E_{\bf k}^+)^2}-\frac{i\omega_n-\xi_{\bf
k}^-}{(i\omega_n)^2-(E_{\bf k}^-)^2}\right],\nonumber\\
&&{\cal B}_{12}(K)=-\frac{1}{2}\left[\frac{\Delta_\mf}{(i\omega_n)^2-(E_{\bf
k}^+)^2}-\frac{\Delta_\mf}{(i\omega_n)^2-(E_{\bf
k}^-)^2}\right],\nonumber\\
&&{\cal B}_{21}(K)=-{\cal B}_{12}(K).
\end{eqnarray}
Here $E_\bk^\pm=\sqrt{(\xi_\bk^\pm)^2+\D_\mf^2}$ with $\x_\bk^\pm=\xi_\bk\pm\l|\bk|$ is the fermion dispersion relation. One can verify that \eqs{mfs1}{mfs4}
satisfy \eqs{g1}{g6}.

Then from the standard Green's function method, the coupled gap and density equations are expressed as
\begin{eqnarray}
\label{gapbcs}
\D_\mf&=&-\frac{U}{2\b V}\Tr\sum_Ki\s_y\cf_\mf(K)\non
&=&-\frac{U\D_\mf}{2 V}\sum_{\a=\pm}\sum_\bk\frac{1-2 n_F(E^\a_\bk)}{2E_\bk^\a},\\
\label{numbcs}
n&=&\frac{1}{\b V}\Tr\sum_K e^{i\w\o_n}\cg_\mf(K)\non
&=&\frac{1}{V}\sum_{\a=\pm}\sum_\bk \ls(u_\bk^\a)^2 n_F(E_\bk^\a)+(v_\bk^\a)^2 n_F(-E_\bk^\a)\rs,\non
\end{eqnarray}
where $n_F(x)=1/[\exp{(\b x)}+1]$ is the Fermi-Dirac function, $e^{i\w\o_n}$ with $\w\ra0^+$ is a convergence factor for the Matsubara
summation, and $(u_\bk^\a)^2=\frac{1}{2}(1+\xi_\bk^\a/E^\a_\bk)$ and $(v_\bk^\a)^2=\frac{1}{2}(1-\xi_\bk^\a/E^\a_\bk)$ are the
Bogoliubov coefficients.

The difference between $\cg_0^{-1}(K)$ and $\cg_\mf^{-1}(K)$ defines the mean-field self-energy,
\begin{eqnarray}
\label{sebcs}
\S_\mf(K)&=&\cg_0^{-1}(K)-\cg_\mf^{-1}(K)\non
&=&-\D_\mf^2i\s_y\tilde{\cg}_0(K)i\s_y.
\end{eqnarray}
If we define a $T$ matrix in the following form,
\begin{eqnarray}
\label{tmf}
t_\mf(Q)&=&-\D_\mf^2\d(Q),
\end{eqnarray}
where $Q=(q_0\equiv i\o_\n, \bq)$ with $\o_\n=2\n\p T$ ($\n$ integer) being the boson Matsubara frequency and $\d(Q)=\b\d_{\n,0}\d^{(3)}(\bq)$,
\eq{sebcs} can be rewritten in a manner of
\begin{eqnarray}
\label{sebcs2}
\S_\mf(K)&=&\frac{1}{\b V}\sum_Q t_\mf(Q)i\s_y\tilde{\cg}_0(K-Q)i\s_y.
\end{eqnarray}
This shows that in the BCS theory, the fermion-fermion pairs contribute to the fermion self-energy only through their condensate at zero momentum,
and these {\it condensed} pairs are associated with a $T$-matrix or propagator (\ref{tmf}).

Furthermore, if we define the mean-field pair susceptibility as
\begin{eqnarray}
\label{cmf}
\c_\mf(Q)&=&\frac{1}{2\b V}\Tr\sum_K\ls \cg_\mf(K)i\s_y\tilde{\cg}_0(K-Q)i\s_y\rs,\non
\end{eqnarray}
we can rewrite the gap equation in the superfluid phase as
\begin{eqnarray}
\label{thoulessmf}
1+U\c_\mf(0)=0,\;\;\; T\leq T_c.
\end{eqnarray}
This suggests that if one considers the {\it uncondensed} pair
propagator or $T$ matrix to be of the form
\begin{eqnarray}
\label{pmf} t_{\rm pair}=\frac{U}{1+U\c_\mf(Q)},\; \; Q\neq0,
\end{eqnarray}
then the gap equation is given by $t_{\rm pair}^{-1}(0)=0$.

It is well known that the critical temperature $T_c$ in the BCS
theory is related to the appearance of a singularity in a $T$ matrix
in the form of \eq{pmf} but with $\D_\mf=0$. This is the so-called
Thouless criterion for $T_c$~\cite{Thouless}. But the meaning of
\eq{thoulessmf} is more general as stressed by Kadanoff and
Martin~\cite{km}. It states that under an asymmetric choice of
$\c_\mf$, the gap equation is equivalent to the requirement that the $T$
matrix associated with the uncondensed pairs remains singular at zero
momentum and frequency for all temperatures below $T_c$.

Although the construction of the uncondensed pair propagator
(\ref{pmf}) in BCS scheme is quite natural, the uncondensed pair has
no feedback to the fermion self-energy (\ref{sebcs2}). 
In the BCS limit (both $|U|$ and $\l$ are small), such a feedback may not be important, but if
the system is strongly coupled (for large $|U|$ and/or for large $\l$ for EO or S SOC), this feedback will be essential.
The simplest way to include the feedback effects is to replace
$t_\mf$ in \eq{sebcs2} by $t_\mf+t_{\rm pair}$. But to make such an
inclusion self-consistent, $t_{\rm pair}$ should be somewhat
modified which we now discuss.

\subsection {$G_0G$ Formalism at $T\leq T_c$}\label{below}
The BCS theory involves the contribution to the self-energy from the condensed pairs only, but
generally, in superfluid phase, the self-energy consists of two distinctive contributions, one
from the superfluid condensate, and the other from thermal or quantum pair fluctuations. Correspondingly,
it is natural to decompose the self-energy into two additive terms,
\begin{eqnarray}
\S(K)&=&\frac{1}{\b V}\sum_Q t(Q)i\s_y\tilde{\cg}_0(K-Q)i\s_y\non&=&\S_\sf(K)+\S_\pg(K),
\end{eqnarray}
with the $T$ matrix accordingly given by
\begin{eqnarray}
\label{tpg0} t(Q)&=&t_\sf(Q)+t_\pg(Q),\non
t_\sf(Q)&=&-\D_\sf^2\d(Q),\non
t_\pg(Q)&=&\frac{U}{1+U\c(Q)},\;\; Q\neq0,
\end{eqnarray}
where the subscript $\sf$ and $\pg$ indicate that these terms are responsible to the
superfluid condensate and pseudogap in fermionic dispersion relation.
See \fig{feynman} for the Feynman diagrams for $t_\pg(Q)$ and
$\S(K)$. Comparing with the BCS scheme, $t_\mf(Q)$ in \eq{sebcs2} is
replaced by $t(Q)$, and $\S(K)$ now contains the feedback of
uncondensed pairs. Inspired by
\eq{cmf}, we now choose the pair susceptibility $\c(Q)$ to be the following asymmetric $G_0G$ form,
\begin{eqnarray}
\label{suscep}
\c(Q)&=&\frac{1}{2\b V}\sum_K\cg(K)i\s_y\tilde{\cg}_0(K-Q)i\s_y.
\end{eqnarray}

In spirit of Kadanoff and Martin, we now propose the superfluid
instability condition or gap equation as [extension of
\eq{thoulessmf}]
\begin{eqnarray}
\label{thouless} 1+U\c(0)=0,\;\; T\leq T_c.
\end{eqnarray}
We stress here that this condition has quite clear physical meaning
in BEC regime. The dispersion relation of the bound pair is given by
$t_\pg^{-1}(Q)=0$, hence $t_\pg^{-1}(0)\propto\m_b$ with $\m_b$ the
effective chemical potential of the pairs. Then the BEC condition
requires $\m_b=0$, and thus $t_\pg^{-1}(Q)=0$,  for all $T\leq T_c$.

The gap equation (\ref{thouless}) tells us that $t_\pg(Q)$ is highly
peaked around $Q=0$, so we can approximate $\S_\pg$ as
\begin{eqnarray}
\label{spg} \S_\pg(K)\simeq -\D_\pg^2i\s_y\tilde{\cg}_0(K)i\s_y,\;\;T\leq T_c,
\end{eqnarray}
where we have defined the pseudogap parameter $\D_\pg$ via
\begin{eqnarray}
\label{pseudogap} \D^2_\pg=-\frac{1}{\b V}\sum_{Q\neq0} t_\pg(Q).
\end{eqnarray}
The total self-energy now is written in a BCS-type form
\begin{eqnarray}
\label{sfull} \S(K)= -\D^2i\s_y\tilde{\cg}_0(K)i\s_y,
\end{eqnarray}
but with $\D^2=\D^2_\sf+\D^2_\pg$.
It is clear that $\D_\pg$ also
contributes to the energy gap in fermionic excitation.
Physically, the pseudogap $\D_\pg$ below $T_c$ can be interpreted as
extra contribution to the excitation gap of fermion: an additional energy is needed to overcome the
residual binding between fermions in a thermal excited pair to
produce fermion-like quasi-particles. One should note that
$\D_\pg$ is associated with the thermal fluctuation of the pairs
$\D^2_\pg(T)\sim\lan\D^2(T)\ran-\lan\D(T)\ran^2$~\cite{prb61:11662,Chen:1998zz}
hence it does not lead to superfluid (symmetry breaking). Besides, at $T=0$, the $G_0G$ formalism recovers
the BCS theory, hence the $G_0G$ formalism does not involve quantum fluctuation. We note here
that at strong interacting regime the quantum fluctuations could have sizable contributions
to certain quantities like the excitation gap. For example,
at the unitarity and at $T=0$,
the $G_0G$ approach gives $\D\approx 0.69\ve_{\rm F}$ (see, for example, Sec. \ref{pseud}) while quantum Monte Carlo simulation
gives $\D\approx 0.54\ve_{\rm F}$~\cite{carlson}. So neglecting the quantum fluctuations in the $G_0G$ formalism gives
roughly a $30\%$ inaccuracy at $T=0$ for the superfluid gap near unitarity.
\begin{figure}[!htb]
\begin{center}
\includegraphics[width=6.5cm]{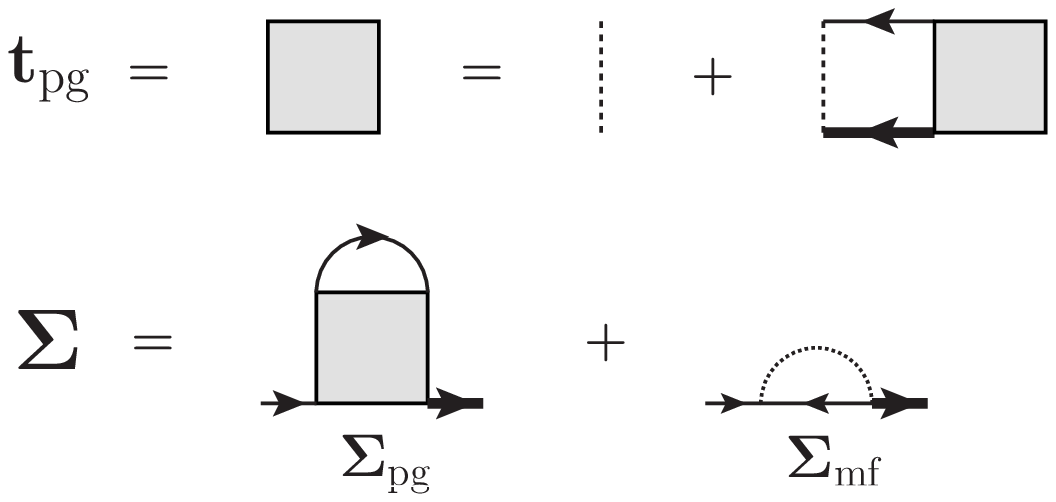}
\caption{Feynman diagrams for the $T$ matrix of non-condensed pairs
and the fermion self-energy in the $G_0G$ formalism.}
\label{feynman}
\end{center}
\end{figure}

With the self-energy (\ref{sfull}), it is easy to see that the gap equation and number equation remain
the forms of \eqs{gapbcs}{numbcs} except the replacement of $\D_\mf\ra\D$:
\begin{eqnarray}
\label{gapbelow}
1&=&-\frac{U}{2 V}\sum_{\a=\pm}\sum_\bk\frac{1-2 n_F(E^\a_\bk)}{2E_\bk^\a},\\
\label{numbelow}
n&=&\frac{1}{\b V}\Tr\sum_K e^{i\w\o_n}\cg(K)\non
&=&\frac{1}{V}\sum_{\a=\pm}\sum_\bk \ls(u_\bk^\a)^2 n_F(E_\bk^\a)+(v_\bk^\a)^2 n_F(-E_\bk^\a)\rs.\non
\end{eqnarray}

The pair susceptibility (\ref{suscep}) can be calculated as,
\begin{eqnarray}
\label{cpg}
\c(Q)&=&\frac{1}{4V}\sum_{\a,\g,s=\pm}\sum_\bk\frac{1}{2}\lb1+s\frac{\x_\bk^\a}{E_\bk^\a}\rb(1-\a\g \ct_{\bk\bq})\non
&&\times\frac{n_F(sE^\a_\bk)-n_F(-\x^\g_{\bq-\bk})}{q_0-sE^\a_\bk-\x^\g_{\bq-\bk}},
\end{eqnarray}
with $q_0=i\o_\n$ and
\begin{eqnarray}
\ct_{\bk\bq}=\sum_{i=1}^3\frac{\l_i^2k_i(q_i-k_i)}{\l^2|\bk||\bq-\bk|}.
\end{eqnarray}
Furthermore, the gap equation (\ref{thouless}) suggests that we can
make the following Taylor expansion for $\c(Q)$,
\begin{eqnarray}
\c(Q)=\c(0)+Z\lb q_0-\sum_{i=1}^3 \frac{1}{2m_{bi}}q_i^2\rb+\cdots,
\end{eqnarray}
where $Z$ is a pair wave-function renormalization factor and $m_{bi}$ is the effective ``boson" mass parameter in the $i$ direction.
A straightforward calculation leads to
\begin{eqnarray}
\c(0)&=&\frac{-1}{2V}\sum_{\a,s=\pm}\sum_\bk\frac{s}{2E_\bk^\a}n_F(sE_\bk^\a),\\
Z&=&\frac{\pt\c(Q)}{\pt q_0}\Big|_{Q=0}\non
&=&\frac{1}{2V}\sum_{\a,s=\pm}\sum_\bk\frac{s}{2E_\bk^\a}\frac{n_F(E_\bk^\a)-n_F(s\x_\bk^\a)}{E^\a_\bk-s\x_\bk^\a},
\end{eqnarray}
\begin{widetext}\begin{eqnarray}
\frac{Z}{2m_{bi}}&=&-\frac{1}{2}\frac{\pt^2\c(Q)}{\pt q_i^2}\Big|_{Q=0}\non
&=&-\frac{1}{8V}\sum_{\a,s=\pm}\sum_\bk\bigg\{\frac{s}{2E_\bk^a}\frac{\l_i^2}{\l^2\bk^2}\lb1-
\frac{\l_i^2k_i^2}{\l^2\bk^2}\rb n_F(sE_\bk^\a)-\frac{1}{2}\lb1+s\frac{\x_\bk^\a}{E_\bk^\a}\rb
\frac{\l_i^2}{\l^2\bk^2}\lb1-\frac{\l_i^2k_i^2}{\l^2\bk^2}\rb\frac{n_F(sE_\bk)-n_F(-\x_\bk^{-\a})}{\x_\bk^{-\a}+sE_\bk^\a}\non
&-&\frac{2s}{E_\bk^\a}\lb\frac{k_i}{m}+\a\frac{\l_i^2k_i}{\l|\bk|}\rb^2
\ls\frac{n_F(sE_\bk^\a)-n_F(-\x_\bk^\a)}{(sE_\bk^\a+\x_\bk^\a)^2}
+\frac{\b n_F(\x_\bk^\a)n_F(-\x_\bk^\a)}{sE_\bk^\a+\x_\bk^\a}\rs\non
&+&\frac{s}{E_\bk^\a}\lb\frac{1}{m}+\a\frac{\l_i^2}{\l|\bk|}-\a\frac{\l_i^4k_i^2}{\l^3|\bk|^3}\rb
\frac{n_F(sE_\bk^\a)-n_F(-\x_\bk^\a)}{sE_\bk^\a+\x_\bk^\a}\bigg\}.
\end{eqnarray}\end{widetext}
According to this Taylor expansion, we apply the
pole approximation to the pair propagator or $T$
matrix $t_\pg(Q)$,
\begin{eqnarray}
\label{tpg1} t_\pg(Q)\simeq\frac{Z^{-1}}{q_0-\sum_{i=1}^3q_i^2/(2m_{bi})}.
\end{eqnarray}
We stress here that in general, the small $Q$ expansion of $t^{-1}_\pg(Q)$
should contain a term $\propto q_0^2$. Without this term, \eq{tpg1} does not respect the
particle-hole symmetry and thus can work, in principle, only when the system becomes bosonic.
At the BCS limit, the system possesses a sharp Fermi surface, the pair propagator should asymptotically
recover the particle-hole symmetry, \ie, the $q_0^2$
term should be kept. However, since at BCS limit the pseudogap
is expected to be very small, applying \eq{tpg1} does not bring much quantitative difference.
Therefore, we will apply \eq{tpg1} to the whole crossover region. We also note that the pole
approximation to the pair propagator generally strengthens the uncondensed pairing and thus leads to amplification of the pseudogap effects, however, it remains
a very good approximation for our qualitative and semi-quantitative analysis.

Substituting the number density equation, the parameter $Z$ can be expressed as
\begin{eqnarray}
Z&=&\frac{1}{\D^2}\ls\frac{n}{2}-\frac{1}{2V}\sum_{\a=\pm}\sum_\bk n_F(\xi_\bk^\a)\rs.
\end{eqnarray}
The expression in the square bracket of the right-hand side is
nothing but the density of the pairs $n_b$, we thus have
$n_b=Z\D^2$.

Substituting \eq{tpg1} into \eq{pseudogap} leads to
\begin{eqnarray}
\label{pseudogap1}
\D^2_\pg&=&\frac{1}{ZV}\sum_\bq
n_B[\sum_{i=1}^3q_i^2/(2m_{bi})]\non
&=&\frac{1}{Z}\prod_{i=1}^3\sqrt{\frac{Tm_{bi}}{2\p}}\z\lb3\over2\rb,
\end{eqnarray}
where $n_B(x)=1/[\exp{(\b x)}-1]$ is the Bose-Einstein function and
a vacuum term was regularized out. It should be stressed that at
zero temperature $\D^2_\pg=0$, hence the $G_0G$ scheme yields the
BCS ground state. It is also worth noting that $\D^2_\pg=n_b^{\rm
uncondensed}/Z$, and hence $\D^2_\sf=n_b^{\rm condensed}/Z$.

Now, \eq{gapbelow}, \eq{numbelow}, as well as \eq{pseudogap1} are
coupled to determine the total excitation gap $\D$, the pseudogap
$\D_\pg$ and the chemical potential $\m$ at a given temperature below $T_c$,
and $T_c$ itself is determined by the vanishing of $\D_\sf$.

\subsection {$G_0G$ formalism at $T\gtrsim T_c$}\label{above}
Above $T_c$, \eq{thouless} does not apply, hence \eq{spg} no longer
holds. To proceed, we extend our more precise $T\leq T_c$ equations
to $T>T_c$ in a simplest fashion. We will continue to use
\eq{sfull} to parameterize the self-energy but with $\D=\D_\pg$, and
ignore the finite lifetime effect associated with normal state
pairs. In the absence of the SOC, it was shown that this is a good approximation when
temperature is not very much higher than $T_c$~\cite{phyc1999,phyrept2005}.
The $T$ matrix $t_\pg(Q)$ at small $Q$ can be approximated now as
\begin{eqnarray}
\label{tpg2} t_\pg(Q)\simeq\frac{Z^{-1}}{q_0-\O_\bq},
\end{eqnarray}
where $\O_\bq=\sum_{i=1}^3q_i^2/(2m_{bi})-\m_b$. Since there is no condensation in
normal state, the effective pair chemical potential $\m_b$ is no
longer zero, instead, it should be calculated from
\begin{eqnarray}
\label{gap2}
Z\m_b&\equiv&t^{-1}(0)=\frac{1}{U}+\c(0)\non&=&\frac{1}{U}-\frac{1}{2V}
\sum_{\a,s=\pm}\sum_\bk\frac{s}{2E_\bk^\a}n_F(sE_\bk^\a).
\end{eqnarray}
This is used as the modified gap equation. Similarly, above $T_c$
the pseudogap $\D_\pg$ is determined by
\begin{eqnarray}
\label{pseudogap2} \D^2_\pg&=&\frac{1}{ZV}\sum_\bq
n_B(\O_\bq)\non&=&\frac{1}{Z}\prod_{i=1}^3\sqrt{\frac{Tm_{bi}}{2\p}}{\rm
Li}_{3\over2}\lb e^{\m_b/T}\rb,
\end{eqnarray}
where ${\rm Li}_n(z)$ is the polylogarithm function. Then \eq{gap2},
\eq{pseudogap2} and the number equation which remains unchanged
determine $\D_\pg$, $\m$ and $\m_b$ at $T>T_c$.

At this point, we comment that at $T\gtrsim T_c$, the pseudogap may be closely related to
the contact intensity $\cal C$ which is introduced by Tan~\cite{tan} through the large momentum tail of
the distribution functions, $n_{\ua,\da}(\bk)\ra{\cal C}/\bk^4,\;\bk\ra\infty$, and underlies
a variety of universal thermodynamical relations for the Fermi gases. To see this, we recall
the relation~\cite{pieri,haussmann,hucont} that ${\cal C}=-(m^2/\b V)\sum_{Q} \G_{\rm pair}(Q)$ with $\G_{\rm pair}$ being the
full propagator of the pairs. At $T\gtrsim T_c$, if we approximate $\G_{\rm pair}$ by $t_{\pg}$, we
can roughly estimate the contact intensity as ${\cal C}\sim m^2\D_\pg^2$.
\section {Results and discussions}\label{resul}
With all the equations settled down, we now present the predictions obtained by
solving them numerically.
We will focus on three different types of SOC: EP ($\l_x=\l_y=0, \l_z=\l$), EO ($\l_x=\l_y=\l,\l_z=0$), and S ($\l_x=\l_y=\l_z=\l$).
In all these cases, we regularize the UV divergence in the gap equations by introducing
the $s$-wave scattering length $a$ through
\begin{eqnarray}
\frac{1}{U}=\frac{m}{4\p a}-\frac{1}{V}\sum_\bk\frac{m}{\bk^2}.
\end{eqnarray}

\subsection {Analytical Results in the Molecular BEC limit}\label{molec}
Let us first examine the molecular BEC limit which can be
achieved by either tuning $1/(\kf a)\ra +\infty$ for fixed SOC or
tuning $\l\ra\infty$ for fixed $1/(\kf a)$ for EO and S SOCs. The former case is well studied and
here we are mainly interested in the latter case.
In the molecular BEC limit, we expect $\m<0$ and $|\m|\gg T_c$. For temperature around or
below $T_c$, we can approximate $\x_\bk^\a/T,E_\bk^\a/T\approx\infty$, and all the equations
become temperature independent. In this limit, the gap equation determines the chemical potential
while the number equation determines the gap. By further expanding the equations in powers of $\D/|\m|$
and keeping several leading terms, we obtain some analytical results for various quantities at $T\lesssim T_c$. (Some of them are
already reported in Refs.~\cite{Vyasanakere:2011,Yu:2011,Hu:2011,Iskin:2011a,He:2011,Vyasanakere:2011b}.)
\\
(I) S case. The chemical potential is well given by
\begin{eqnarray}
\m&\approx&-\frac{E_B}{2},
\end{eqnarray}
where $E_B$ is the binding energy determined by the two-body problem \cite{Vyasanakere:2011,He:2011,Vyasanakere:2011b},
\begin{eqnarray}
E_B&=&m\lambda^2+\frac{1}{4m}\lb\frac{1}{a}+\sqrt{\frac{1}{a^2}+4m^2\l^2}\rb^2.
\end{eqnarray}
The effective pair mass $m_b$ coincides with the molecular effective mass determined at the two-body level. We have
\begin{eqnarray}
\frac{2m}{m_b}&\approx&\frac{7}{3}-\frac{4}{3}\lb\frac{E_B-m\l^2}{E_B}\rb^{3/2}-\frac{2m\l^2}{E_B}.
\end{eqnarray}
Other quantities such as $\Delta$ and $Z$ can be evaluated as
\begin{eqnarray}
\D^2&\approx&\frac{32\ve_{\rm F}}{3\p}{\sqrt{\ve_{\rm F}}\over E_B}\lb\frac{E_B-m\lambda^2}{2}\rb^{3/2},\\
Z&\approx&\frac{n}{2\D^2}=\frac{E_B}{8\p}\lb\frac{m}{E_B-m\lambda^2}\rb^{3/2}.
\end{eqnarray}
(II) EO case. The chemical potential is also given by
\begin{eqnarray}
\m&\approx&-\frac{E_B}{2},
\end{eqnarray}
where two-body binding energy $E_B$ is determined by the algebra equation \cite{Vyasanakere:2011,Yu:2011,Hu:2011,Vyasanakere:2011b},
\begin{eqnarray}
\sqrt{\frac{E_B}{m\lambda^2}}-\frac{1}{2}\ln\frac{\sqrt{E_B}+\sqrt{m\lambda^2}}{\sqrt{E_B}-\sqrt{m\lambda^2}}=\frac{m}{\lambda a}.
\end{eqnarray}
The effective pair mass becomes anisotropic and is given by
\begin{eqnarray}
m_b^\perp&\approx&2m\ls1-\frac{m\lambda^2}{2E_B}-\frac{E_B-m\lambda^2}{2E_B}\ln\frac{E_B-m\lambda^2}{E_B}\rs^{-1},\nonumber\\
m_b^\parallel&\approx& 2m.
\end{eqnarray}
Other quantities such as $\Delta$ and $Z$ can be evaluated as
\begin{eqnarray}
\D^2&\approx&\frac{8\ve_{\rm F}(E_B-m\lambda^2)}{3\p}\sqrt{\frac{2\ve_{\rm F}}{E_B}},\\
Z&\approx&\frac{n}{2\D^2}=\frac{\sqrt{m^3E_B}}{8\p (E_B-m\lambda^2)}.
\end{eqnarray}
(III) EP case. We find that the EP case is trivial. Increasing $\lambda$ can not induce a BCS-BEC crossover. For large and positive $1/(k_Fa)$, the
EP SOC only induce a shift for the chemical potential,
\begin{eqnarray}
\m&\approx&-\frac{1}{2ma^2}-\frac{m\l^2}{2}.
\end{eqnarray}
The pair effective mass is almost isotropic and is given by
\begin{eqnarray}
m_b^\perp&\approx&m_b^\parallel\approx 2m.
\end{eqnarray}
Other quantities such as $\Delta$ and $Z$ just recover the usual results without SOC,
\begin{eqnarray}
\D^2&\approx&\frac{8\ve_{\rm F}}{3\p}\frac{\kf}{ma},\\
Z&\approx&\frac{m^2a}{8\p}.
\end{eqnarray}

The critical temperature in the BEC limit, $T_\bec$, is determined by the number equation
\begin{eqnarray}
n_B=\frac{1}{V}\sum_\bk\frac{1}{\exp{\ls\frac{\ve_B(\bk)}{T_\bec}\rs}-1},
\end{eqnarray}
where $\ve_B(\bk)=\sum_{i=1}^3k_i^2/(2m_{bi})$. This leads to $T_\bec=2\p[n_B/(\sqrt{\P_i m_{bi}}\z(3/2))]^{3/2}$ in three dimensions. Setting
$n_B=n/2$, we obtain
\begin{eqnarray}
T_\bec\approx0.218\ve_{\rm F}\prod_{i=1}^3\lb\frac{2m}{m_{bi}}\rb^{1/3}.
\end{eqnarray}
Therefore, in the molecular BEC limit, $T_\bec$ is only a function of the combined dimensionless parameter $\eta=1/(m\lambda a)$.
For S SOC we have
\begin{eqnarray}
m_b&=&\left\{\begin{array}{ll}
\displaystyle 6m, &\displaystyle \eta\ra -\infty\\
\displaystyle 2m, &\displaystyle \eta\ra +\infty \\
\displaystyle 2.32m, &\displaystyle \eta\ra 0
\end{array}\right.\
\end{eqnarray}
and hence
\begin{eqnarray}
\label{tbecs}
T_\bec&=&\left\{\begin{array}{ll}
\displaystyle 0.0726\ve_{\rm F}, &\displaystyle \eta\ra -\infty\\
\displaystyle 0.218\ve_{\rm F}, &\displaystyle \eta\ra +\infty \\
\displaystyle 0.188\ve_{\rm F}, &\displaystyle \eta\ra 0.
\end{array}\right.\
\end{eqnarray}
For EO SOC, we obtain
\begin{eqnarray}
\label{tbeceo}
m_b^\perp&=&\left\{\begin{array}{ll}
\displaystyle 4m, &\displaystyle \eta\ra -\infty\\
\displaystyle 2m, &\displaystyle \eta\ra +\infty \\
\displaystyle 2.40m, &\displaystyle \eta\ra 0
\end{array}\right.\
\end{eqnarray}
and
\begin{eqnarray}
T_\bec&=&\left\{\begin{array}{ll}
\displaystyle 0.137\ve_{\rm F}, &\displaystyle \eta\ra -\infty\\
\displaystyle 0.218\ve_{\rm F}, &\displaystyle \eta\ra +\infty \\
\displaystyle 0.193\ve_{\rm F}, &\displaystyle \eta\ra 0.
\end{array}\right.\
\end{eqnarray}
As we will see, the above obtained $T_\bec$ coincide well with our numerical results in Sec.\ref{tc}.

\subsection {Superfluid Critical Temperature}\label{tc}
\begin{figure}[!htb]
\begin{center}
\includegraphics[width=7cm]{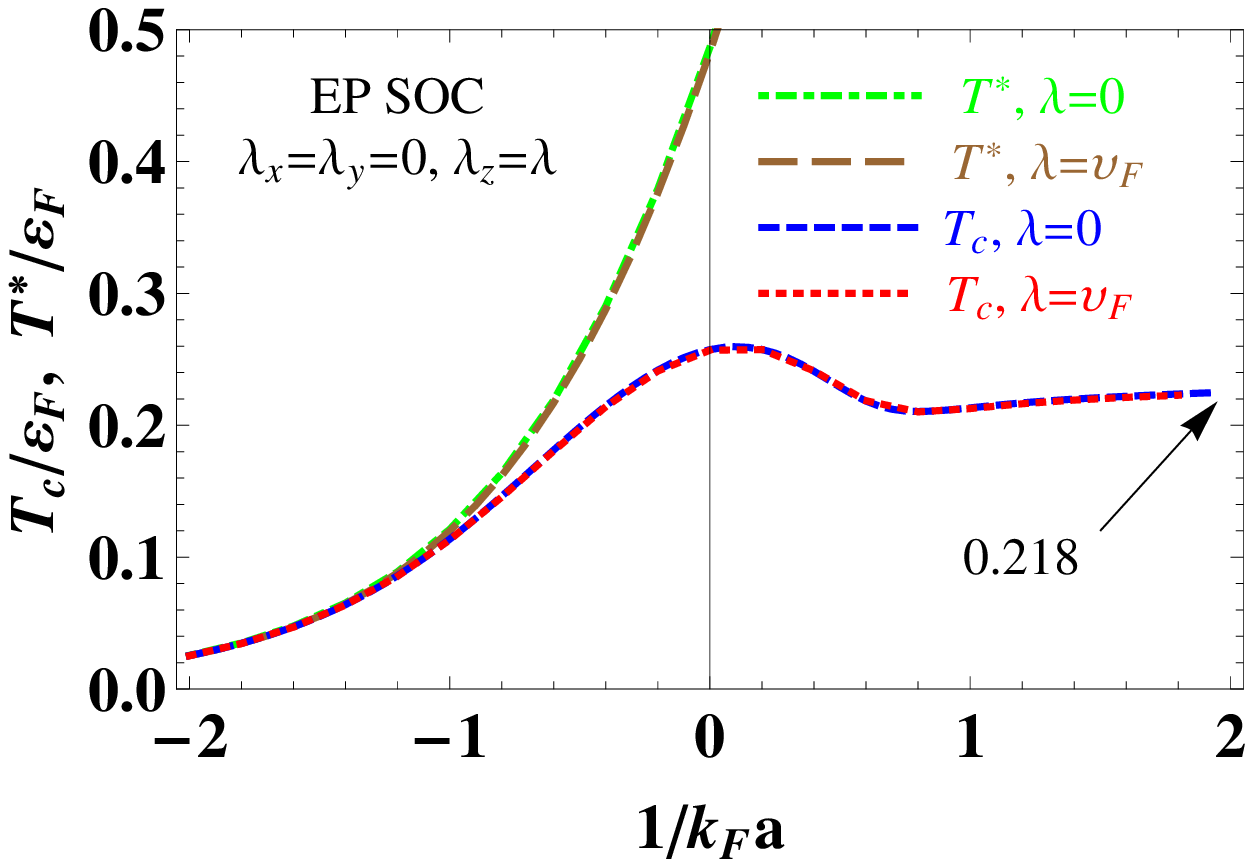}
\includegraphics[width=7cm]{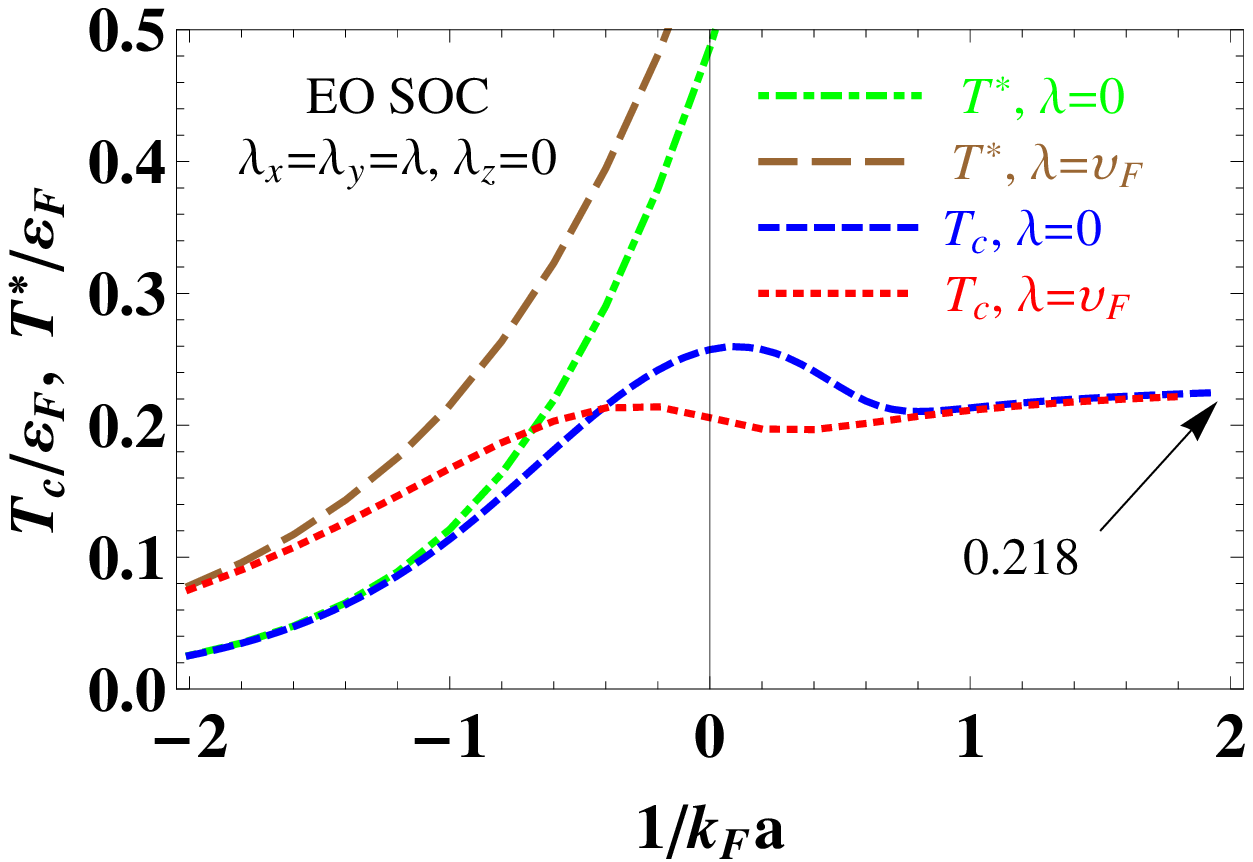}
\includegraphics[width=7cm]{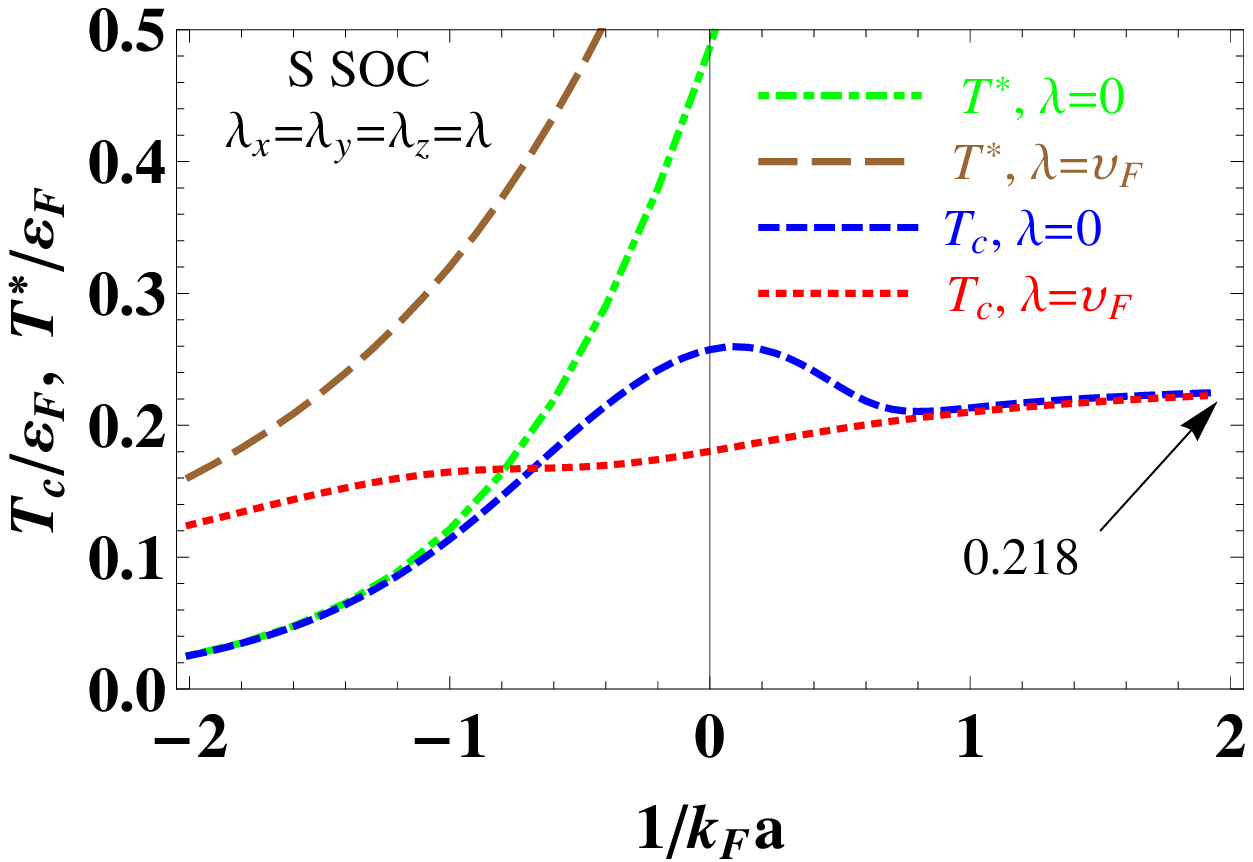}
\caption{(Color online) The critical temperature $T_c$ scaled by the
Fermi energy $\ve_F$ as a function of the gas parameter $1/(\kf a)$ for
fixed SOC $\l=0$ and $\l=\vf$. Also shown is the pair dissociation temperature $T^*$. } \label{tca}
\end{center}
\end{figure}
By numerically solving the set of coupled gap, number density, and pseudogap equations, we can obtain the superfluid order
parameter $\D_\sf$, the pseudogap $\D_\pg$, and the fermion chemical potential $\m$.
The superfluid critical temperature $T_c$ is determined by the vanishing of the superfluid order parameter $\D_\sf$.
The numerical results for $T_c/\ve_{\rm F}$ as a function of the gas parameter $1/(\kf a)$ is shown in \fig{tca}
for $\l=0$ and $\l=\vf$. $T_c/\ve_{\rm F}$ as a function of the SOC $\l$ is
shown in \fig{tcl} for fixed $1/(\kf a)=-2$ and $1/(\kf a)=\infty$. Also shown is the critical temperature predicted
by the BCS theory, $T^*/\ve_{\rm F}$, which is determined by the vanishing of $\D_\mf$. It is monotonously
increasing as $1/(\kf a)$ or $\l/\vf$ increased. The BCS theory loses the pairing-fluctuation effect and
does not give reliable critical temperature particularly at large $1/(\kf a)$ or $\l/\vf$ where $T_c$
is mainly determined by the bosonic degrees of freedom.

For all three types of SOC, we find that $T_c$ is a smooth function of $1/(\kf a)$ and $\l/\vf$, and the superfluid phase transition
is always of second order for the whole crossover region (see next subsection). Also, it can be seen that $T_c$ is not
a monotonous function of $1/(\kf a)$ when $\l$ is small: There is a local maximum in $T_c$ curve around the unitary
point. Similar local maximum also appears when one uses the Nozieres-Schmitt-Rink approach to determine $T_c$ in the absence of SOC. It may be understood by noticing that
the BEC critical temperature is increased when repulsive interactions between bosons are turned on~\cite{Andersson}.
We note that for a Rashba spin-orbit coupled Fermi gas, the superfluid transition temperature has been roughly estimated by approximating the
system as a non-interacting mixture of fermions and rashbons and has been found to increase monotonously across the BCS-BEC crossover~\cite{Yu:2011}.

\begin{figure}[!htb]
\begin{center}
\includegraphics[width=7cm]{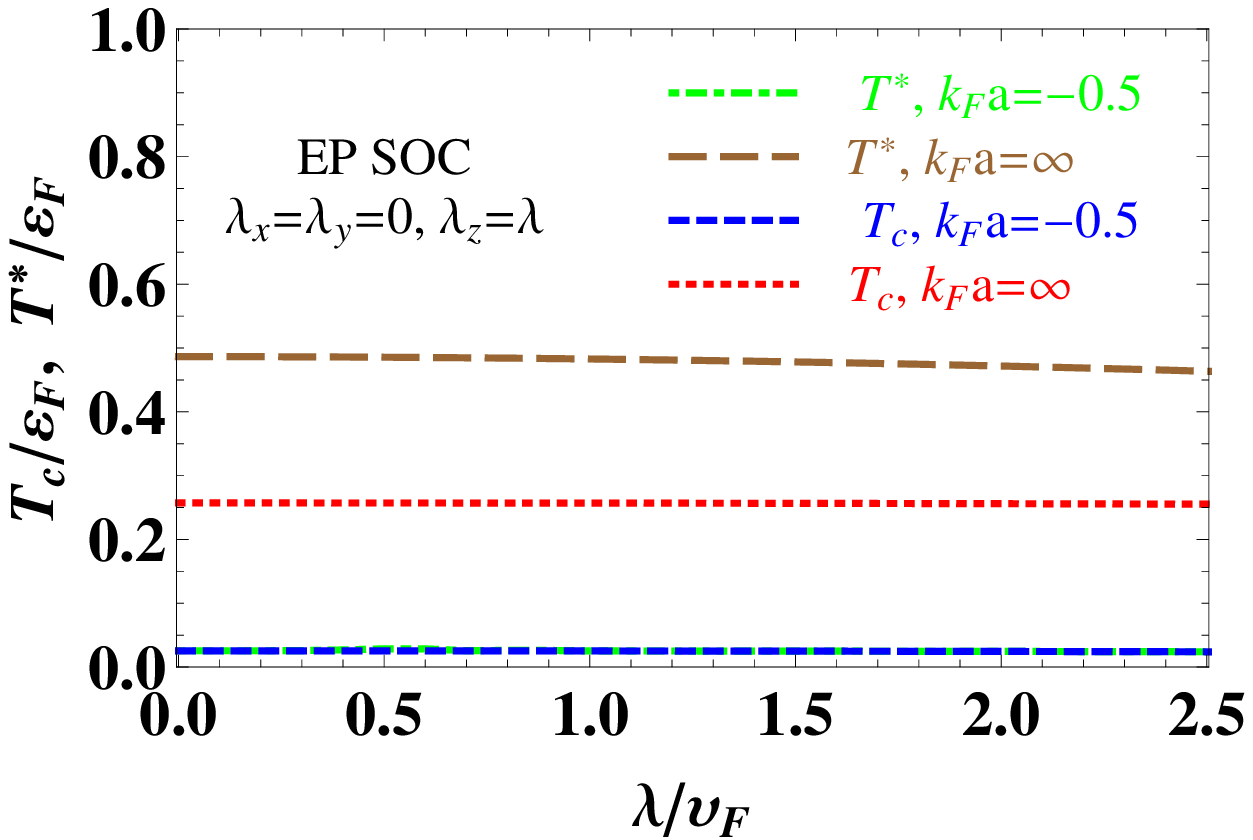}
\includegraphics[width=7cm]{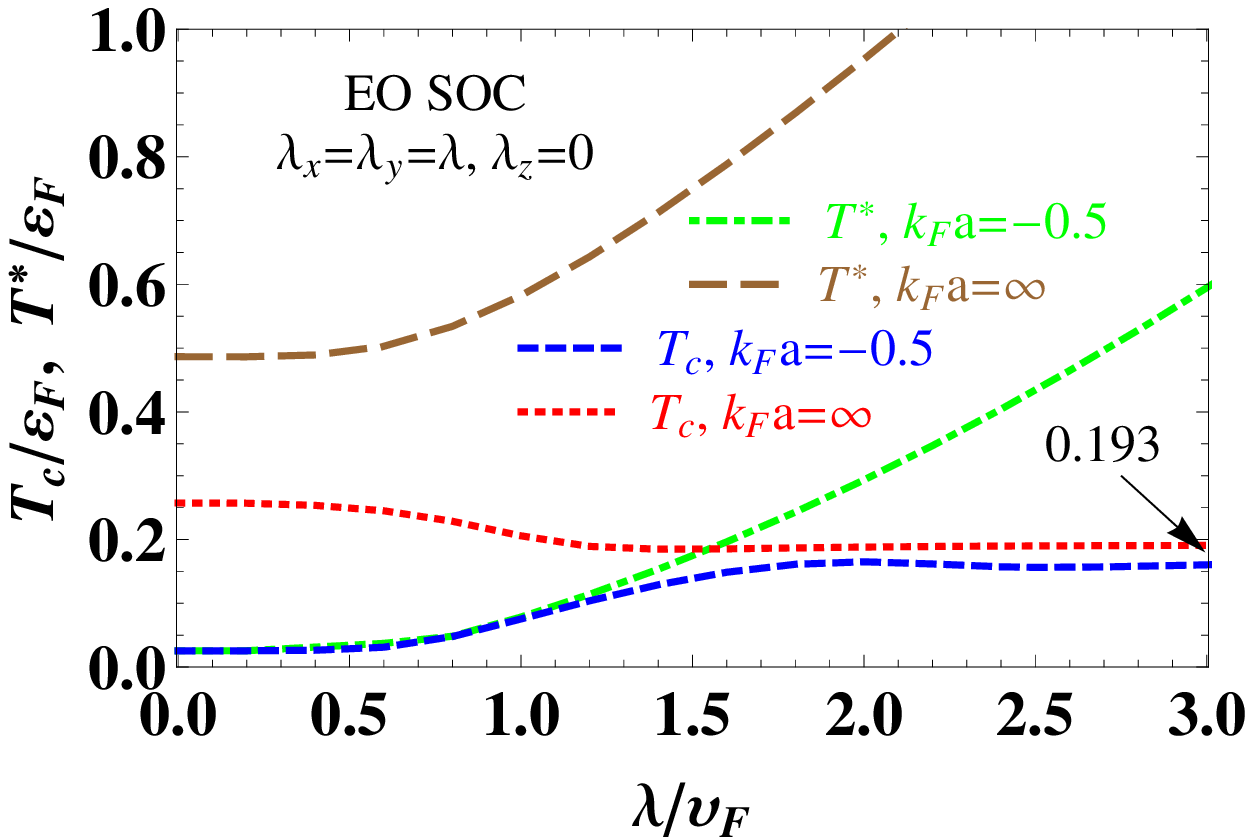}
\includegraphics[width=7cm]{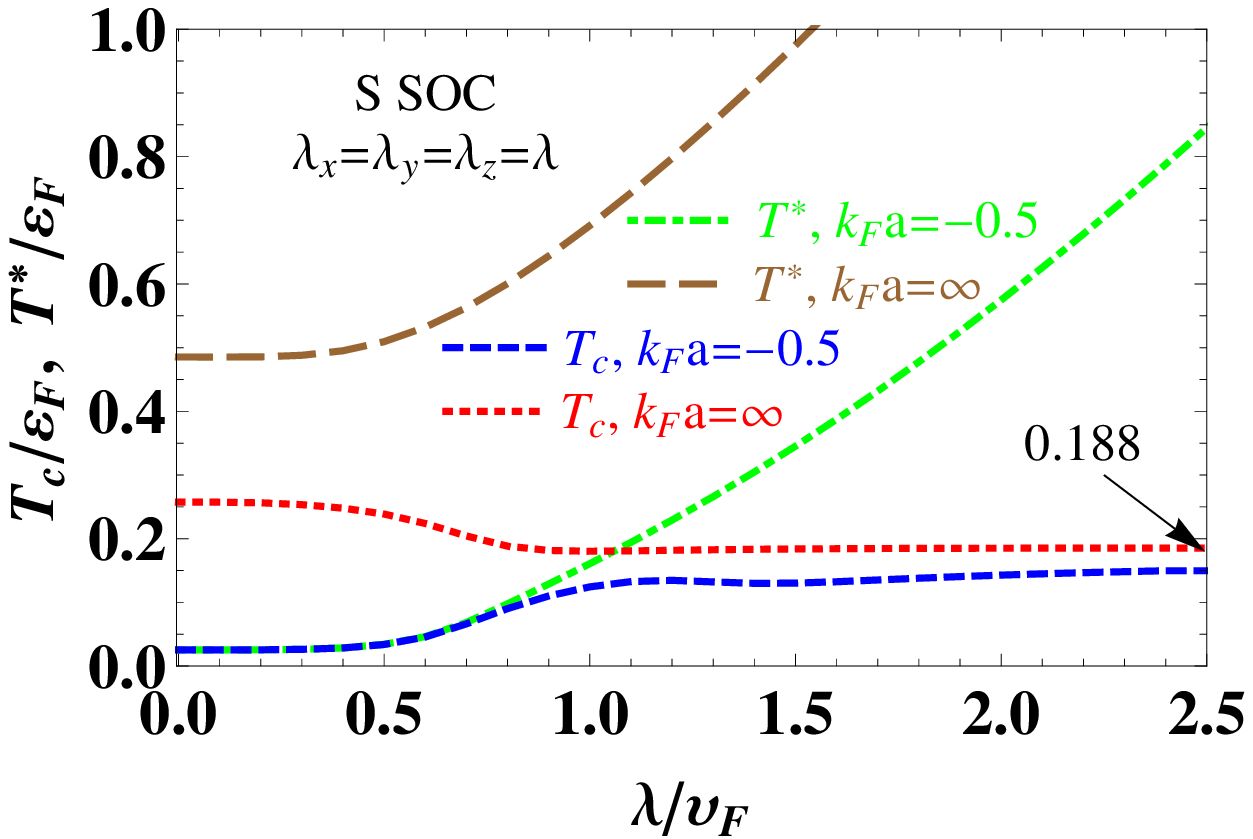}
\caption{(Color online) The critical temperature $T_c$ and the dissociation temperature $T^*$ scaled by the
Fermi energy $\ve_F$ as a function of the SOC $\l/\vf$ for fixed gas parameters
$1/(\kf a)=-2$ and $1/(\kf a)=0$. } \label{tcl}
\end{center}
\end{figure}
From the top panels of \fig{tca} and \fig{tcl} we see that the EP SOC does not affect $T_c$ and $T^*$. This is
consistent with the observation that EP SOC solely does not lead to new novel bound state and the fermion excitation gap does not change~\cite{Vyasanakere:2011b}. This can be understood by noticing
that the EP SOC in Hamiltonian (\ref{lagr}) can be gauged away by using the gauge transformation
$\j_\ua\ra e^{-im\l z}\j_\ua$ and $\j_\da\ra e^{im\l z}\j_\da$, resulting only a constant shift in the chemical potential, $\m\ra \m+m\l^2/2$.

For the EO and S SOCs, we observe from \fig{tca} that, comparing to the case without SOC, the SOC suppresses $T_c$ for $\kf a$ close to unitarity while increases $T_c$ at the BCS regime. To further understand how $T_c$ is influenced by the SOC, we turn to \fig{tcl}. From \fig{tcl} we see that $T_c$ is not
sensitive to $\l$ for $\l\ll\vf$ and $\l\gg\vf$, but becomes sensitive to $\l$
for $\l\sim\vf$: One can identify a BCS-BEC crossover solely induced by $\l$ around $\l\approx\vf$. This coincides with previous zero-temperature studies~\cite{Vyasanakere:2011,Hu:2011,Yu:2011,Iskin:2011a,Gong:2011,Han:2011,Yi:2011,He:2011jva,He:2011,other}.
Although at large negative $1/(\kf a)$,
$T_c$ increases (almost) monotonously as $\l$ grows, near the resonance, we find that $T_c$ is a decreasing
function of $\l$, in contrast Ref.~\cite{Liao:2012} where the authors predicted an increasing $T_c$ along with $\l$. We note that for large enough $\l$, our result converges correctly to a universal molecular limit $T_\bec$
either near the unitarity or at the BCS regime (see Sec. \ref{molec}).

It should be stressed that $T^*$ sets a lower bound for the pair dissociation temperature above which
the pairs essentially dissociate due to thermal excitations. Previous study in the absence of the SOC shows
that it is a good approximation to set $T^*$ on the BCS side or near unitary as the pair dissociation temperature~\cite{phyc1999,
phyrept2005}. So we refer to $T^*$ as the pair dissociation temperature in \fig{tca} and \fig{tcl}.
The region between $T_c$ and $T^*$ is a pseudogap dominated
window in which a normal state is no longer described by the Landau Fermi liquid theory.
\fig{tca} and \fig{tcl} show that even for large negative $1/(\kf a)$
the pseudogap dominated region can be sizable once the SOC is large. Thus, the spin-orbit coupled Fermi gas may provide a new platform to
study the formation of pseudogap in fermionic systems.

\subsection {Pseudogap}\label{pseud}
In this subsection, we focus on S and EO SOCs because EP SOC does not bring qualitatively new
features to the temperature dependence of the pseudogap than the $\l=0$ case.
In \fig{dtl0}-\fig{dtuni}, we plot $\D$, $\D_\sf$, and $\D_\pg$ as well as $\D_\bcs$ (in units of $\ve_{\rm F}$, the same below) as functions of temperature.

\begin{figure}[!htb]
\begin{center}
\includegraphics[width=7cm]{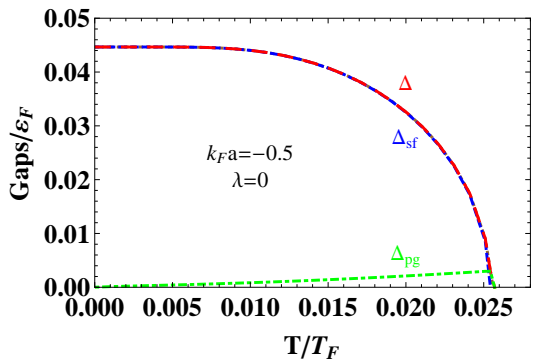}
\caption{(Color online) The temperature dependence of $\D$, $\D_\sf$, and $\D_\pg$
scaled by the Fermi energy for a Fermi gas without SOC at $\kf a=-0.5$.} \label{dtl0}
\end{center}
\end{figure}
\begin{figure}[!htb]
\begin{center}
\includegraphics[width=7cm]{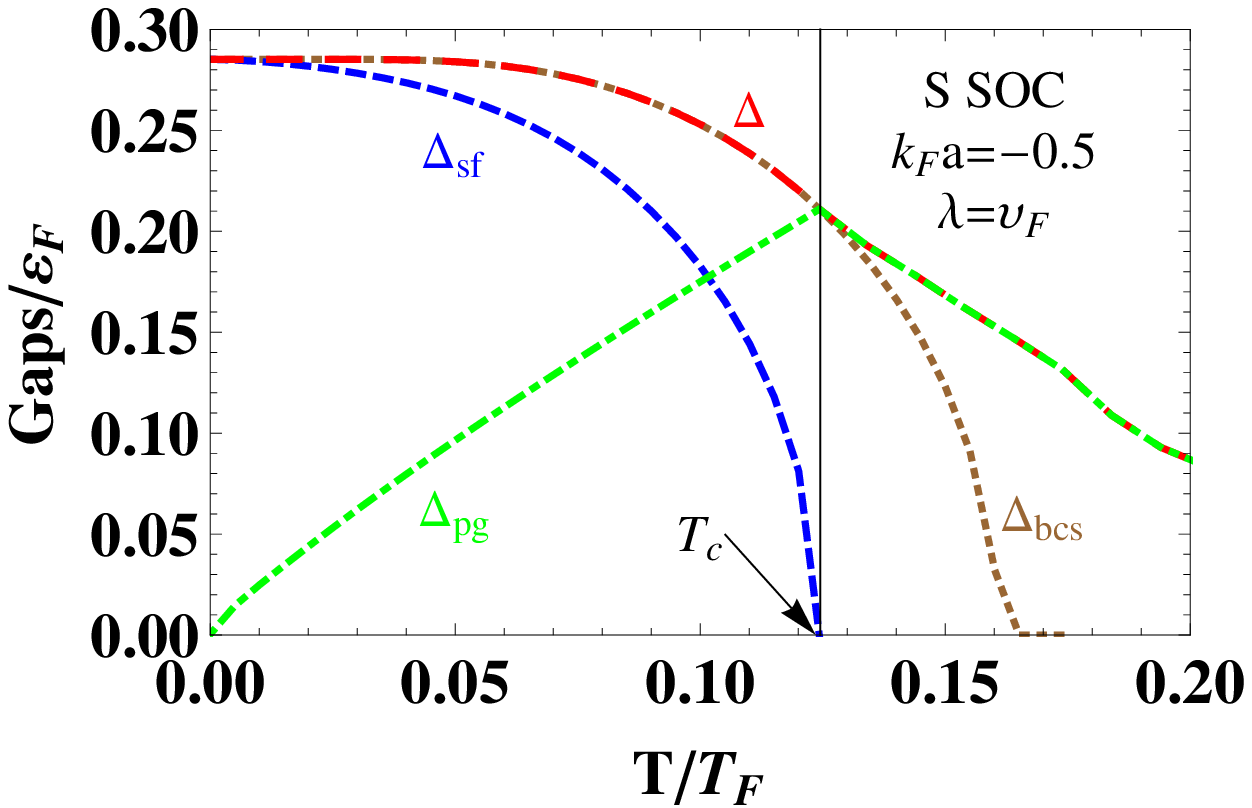}
\includegraphics[width=7cm]{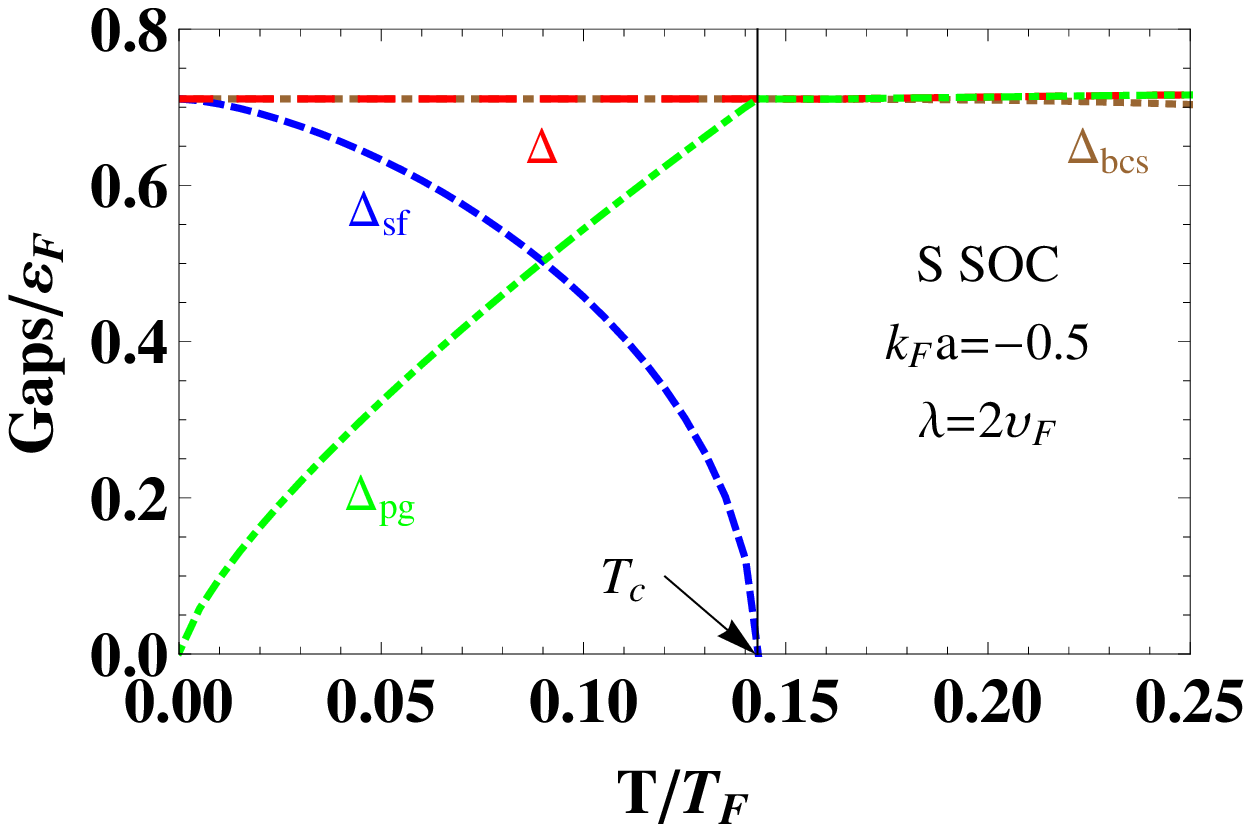}
\caption{(Color online) The temperature dependence of $\D$, $\D_\sf$, and $\D_\pg$ as well as $\D_\bcs$
scaled by the Fermi energy at $\kf a=-0.5$ for S SOC.} \label{dts}
\end{center}
\end{figure}
\begin{figure}[!htb]
\begin{center}
\includegraphics[width=7cm]{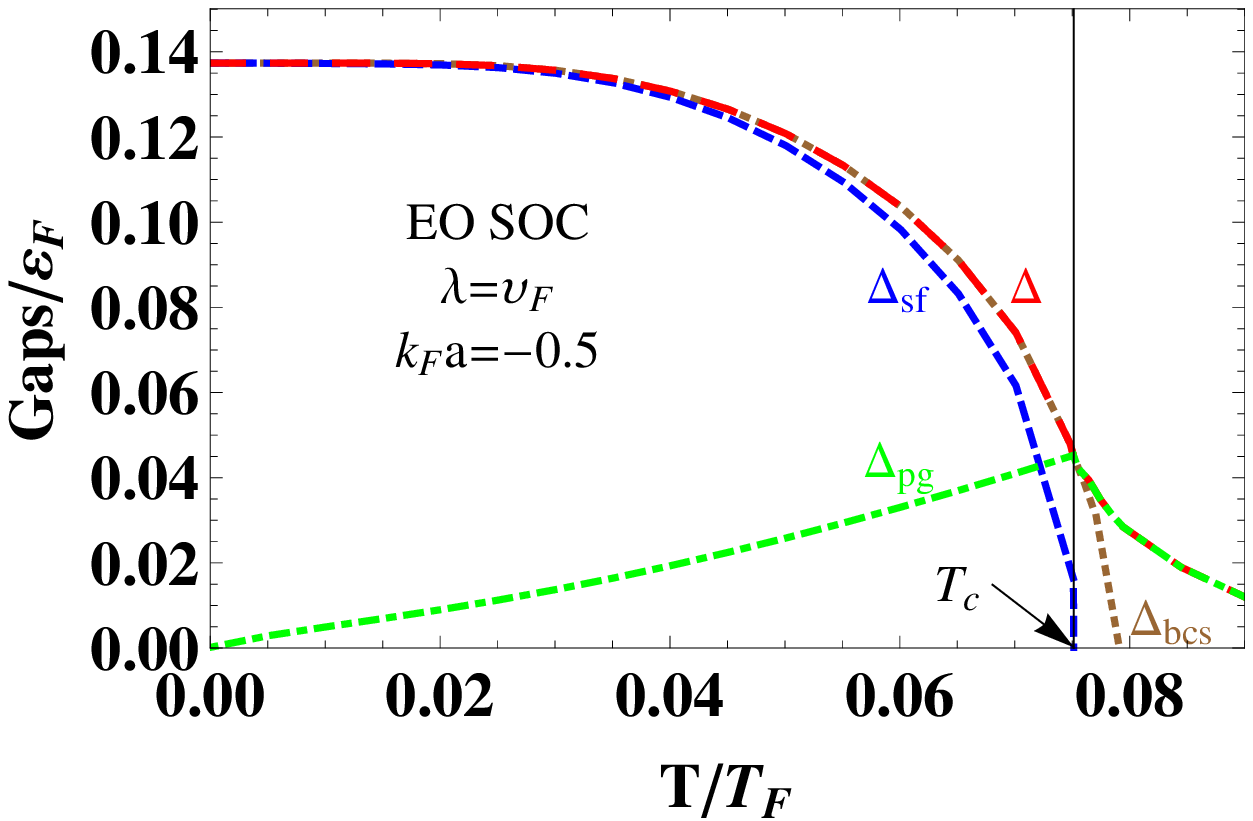}
\includegraphics[width=7cm]{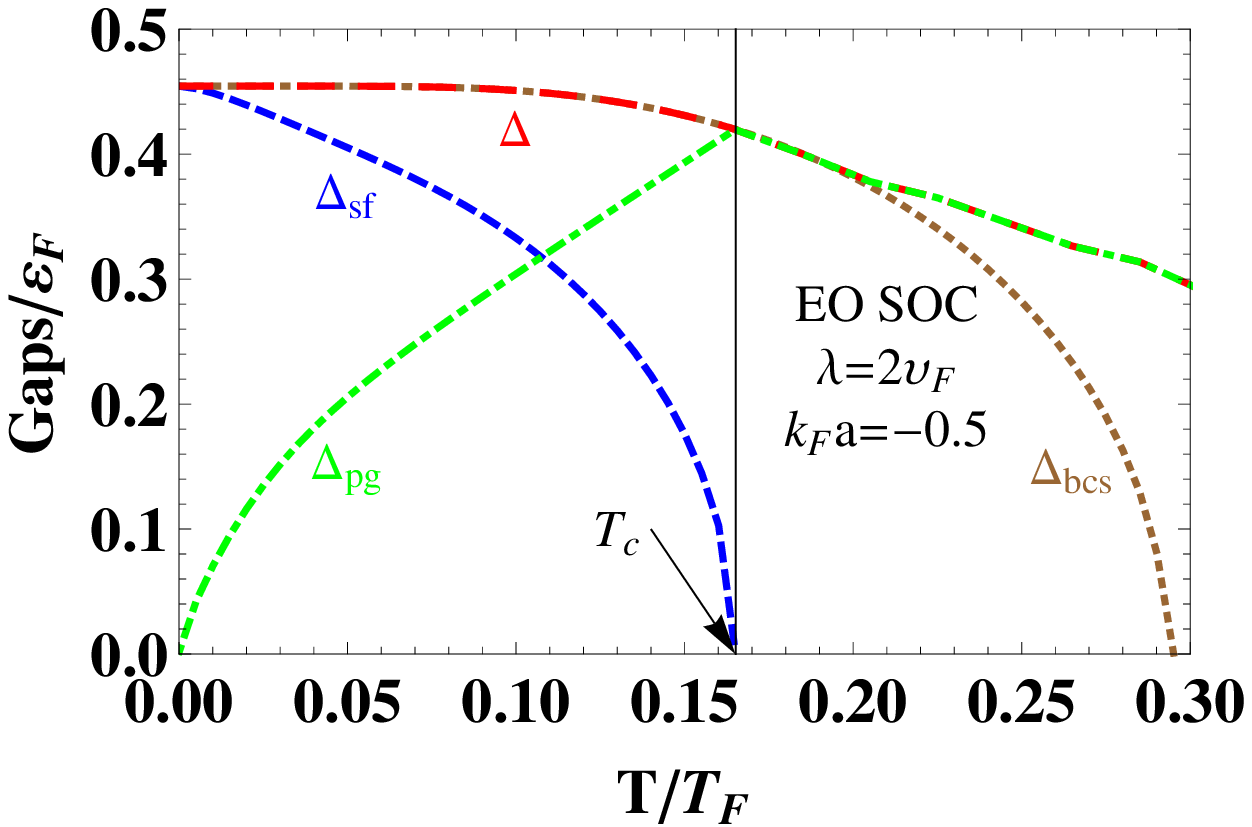}
\caption{(Color online) The temperature dependence of $\D$, $\D_\sf$, $\D_\pg$, and $\D_\bcs$
scaled by the Fermi energy at $\kf a=-0.5$ for EO SOC.} \label{dteo}
\end{center}
\end{figure}
\begin{figure}[!htb]
\begin{center}
\includegraphics[width=7cm]{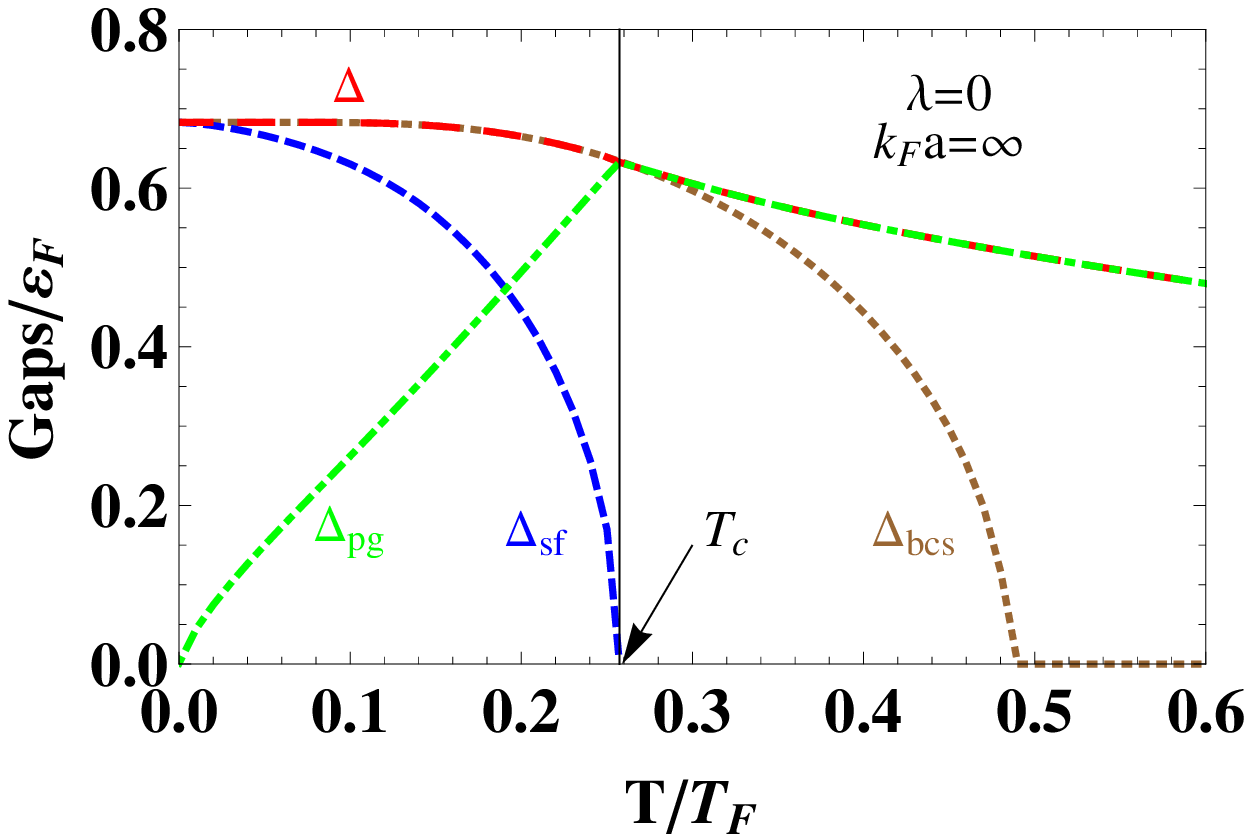}
\includegraphics[width=7cm]{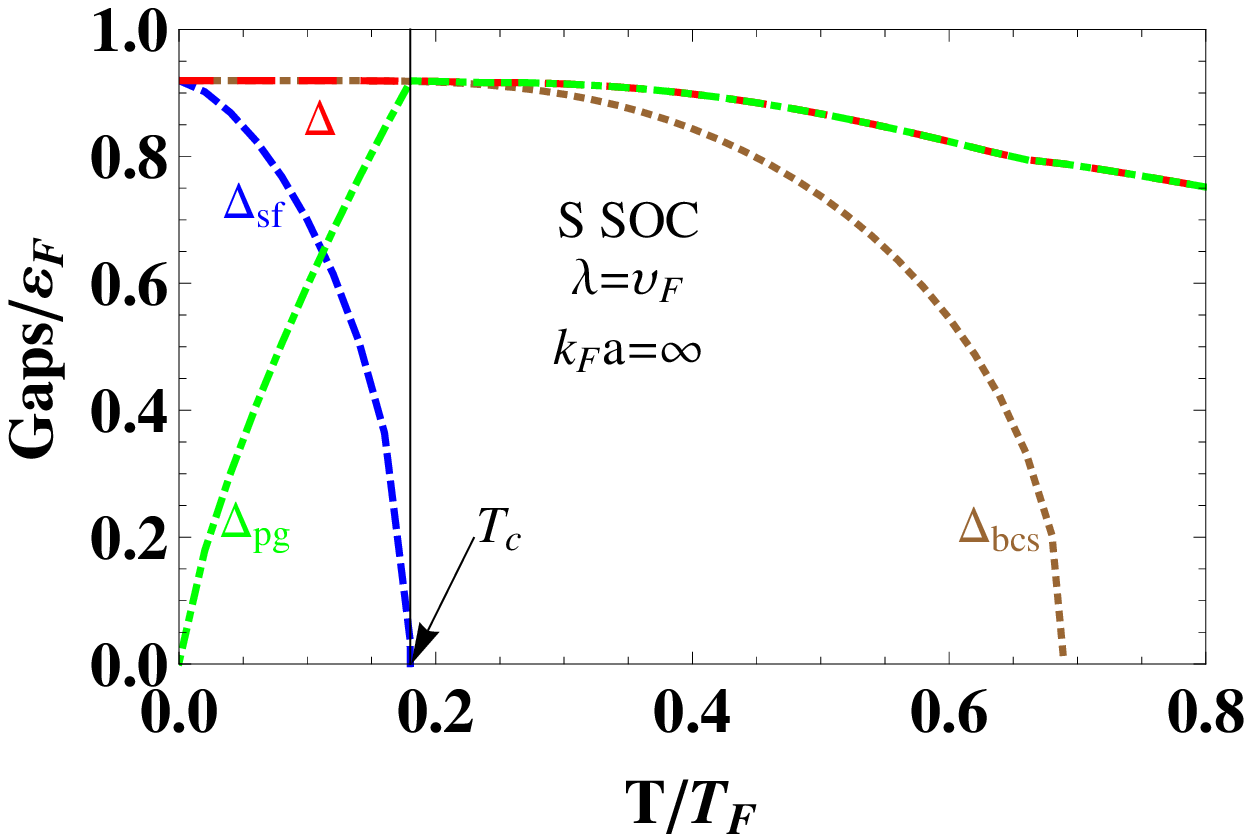}
\includegraphics[width=7cm]{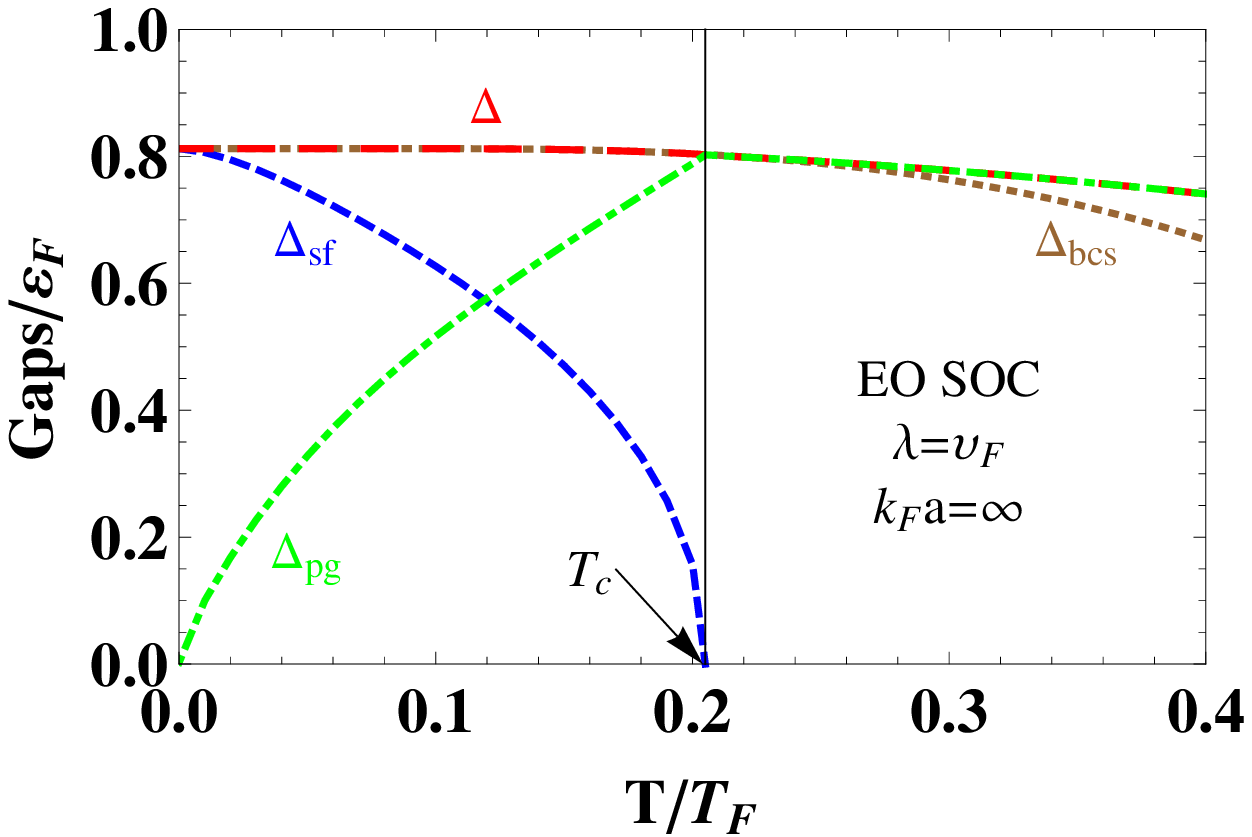}
\caption{(Color online) The temperature dependence of $\D$, $\D_\sf$, $\D_\pg$, and $\D_\bcs$
scaled by the Fermi energy at unitary point for $\l=0$ Fermi gas and for EO- and S-
SOC Fermi gases.} \label{dtuni}
\end{center}
\end{figure}
The common feature for all these figures is that $\D_\sf$ monotonically decrease to zero at $T_c$.
Below $T_c$, $\D_\pg(T)$ is a monotonically increasing function from zero
at $T=0$ where it vanishes according to $\D_\pg\propto T^{3/4}$ (see \eq{pseudogap1}). Above $T_c$, $\D_\pg(T)$ is a monotonically decreasing function from
its maximum value located at $T_c$. This kind of temperature dependence clearly shows that pseudogap is due to the thermally excited pairs:
Below $T_c$ when $T$ goes higher more pairs are excited from the condensate and at $T_c$ all condensed pairs are thermally excited; after that the
thermal motion of the pair participators begins to dissociate the pairs and hence $\D_\pg$ (more precisely, $Z\D_\pg^2$) begins to decrease. Although
the physical pictures are clear, at temperature much higher than $T_c$ our formalism may fail since the finite life-time of the pairs, which is not included in our formalism, may
become important.

By comparing \fig{dts}-\fig{dteo} to \fig{dtl0} and by comparing two bottom panels of \fig{dtuni} to the panel on the top,
one can see that although the SOC does not modify the general tendency of the temperature dependence of the gaps,
large SOC significantly enlarges the pseudogap window in the normal phase.
Such pseudogap window may be detected by RF spectroscopy measurements, which we now turn to study.

\subsection {RF Spectroscopy}\label{rf}
\begin{figure}[!htb]
\begin{center}
\includegraphics[width=7cm]{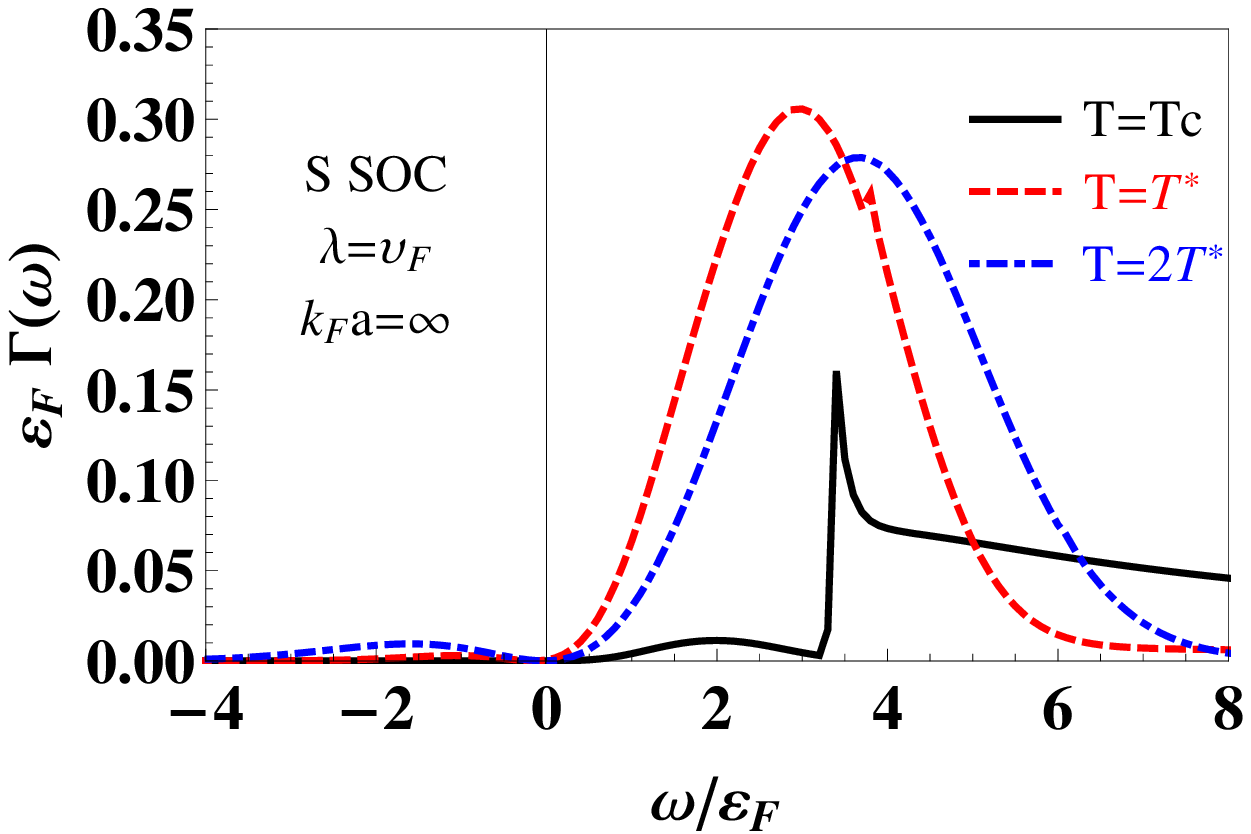}
\includegraphics[width=7cm]{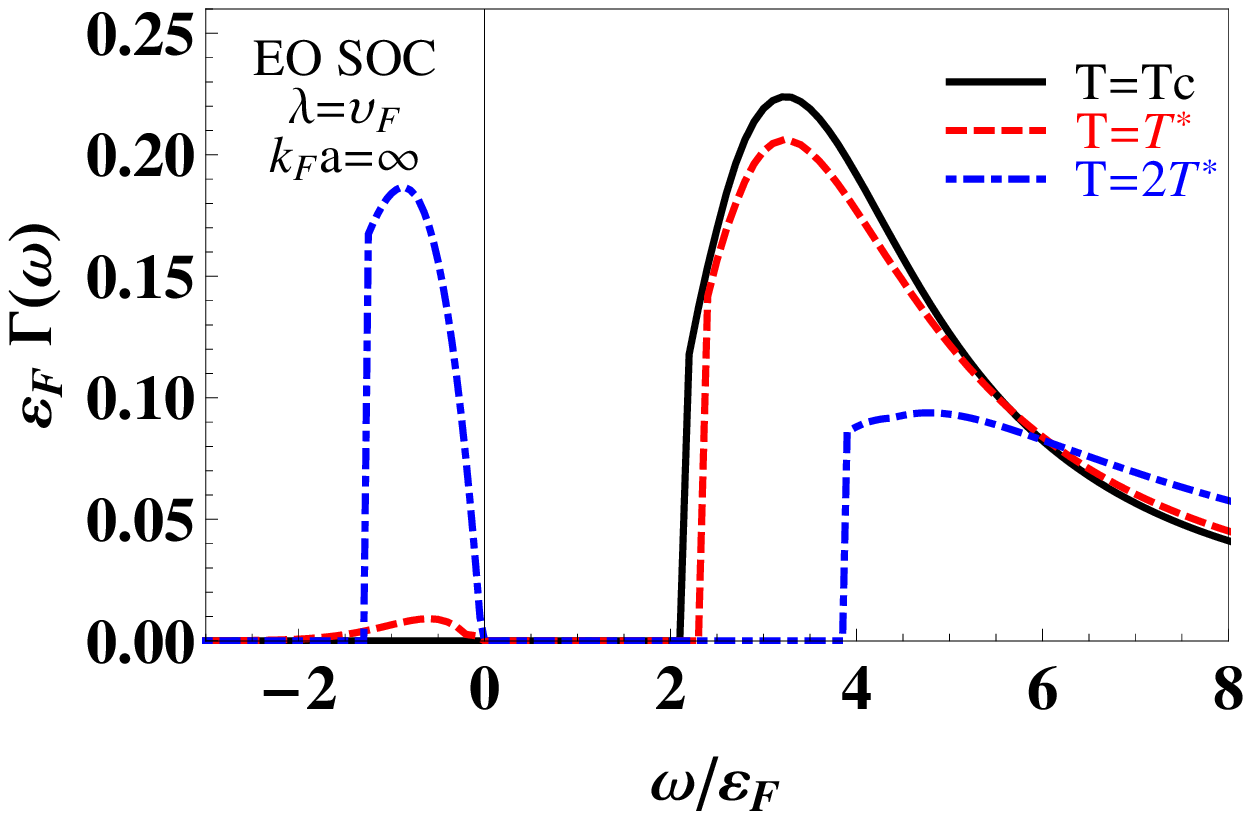}
\includegraphics[width=7cm]{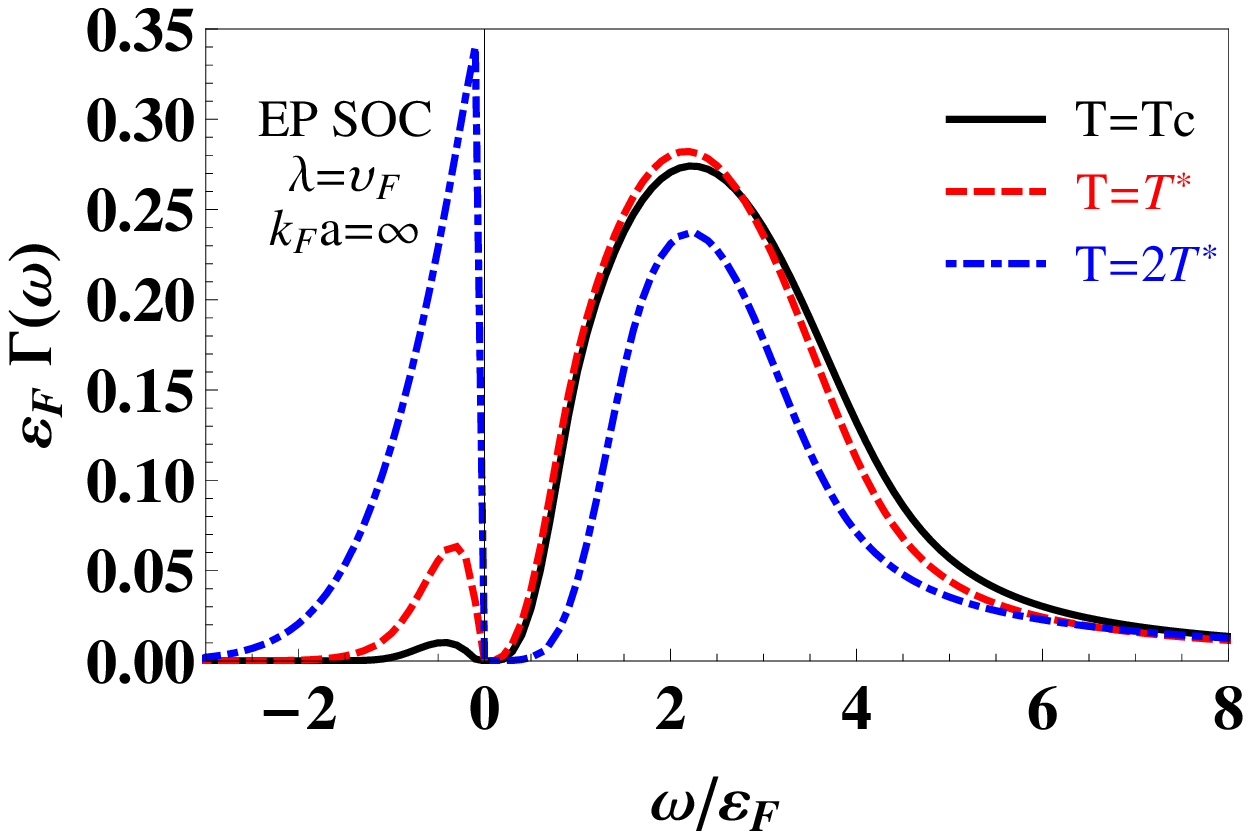}
\caption{(Color online) The RF spectrum $\G(\o)$ in unit of $1/\ve_{\rm F}$ for SO coupled unitary Fermi gases at resonance.} \label{figrf}
\end{center}
\end{figure}
The radio-frequency (RF) spectroscopy has been proven to be very successful in probing the fermionic pairing, quasi-particle excitation spectrum, and
superfluidity. For an atomic Fermi gas with two hyperfine states, $|\uparrow\ran$ and $|\downarrow\ran$,
the RF laser drives transitions between one of the hyperfine states (i.e., $|\da\ran$)
and an empty hyperfine state $|3\ran$ which lies above it by
an energy $\o_{\da 3}$ (which is set to zero because it can be absorbed into the chemical potential)
due to the magnetic field splitting in bare
atomic hyperfine levels. The Hamiltonian for RF-coupling
may be written as,
\begin{eqnarray}
H_{\rm RF}=V_0\int d^3\br \ls \j^\dag_3(\br)\j_\da(\br)+{\rm H.C.}\rs
\end{eqnarray}
where $\j^\dag_3(\br)$ is the field operator which creates an
atom at the position $\br$ and $V_0$ is the strength
of the RF drive and is related to a Rabi frequency $\o_R$ by $V_0=\o_R/2$.

Let us now assume that there is no interaction between the third state and
the spin-up or spin-down states, i.e., there is no final state effect. This approximation sounds valid for $^{40}K$ atoms, where the $s$-wave scattering length between the spin-down state and the third hyperfine state
is small (i.e., $\sim 200$ Bohr radii)~\cite{Bloch:2008}. Within this
approximation and taking into account that the third state is not occupied initially, the transfer strength (integrated RF spectrum)
 per spin-down atom can be written by ($V_0 = 1$),
\begin{eqnarray}
\G(\o)=\frac{1}{Vn_\da}\sum_\bk\ca_{\da\da}(\bk,\x_\bk-\o)n_F(\x_\bk-\o),
\end{eqnarray}
where $n_\da$ is the number density of the spin-down fermion and $\ca_{\da\da}=-(1/\p)\im \cg_{\da\da}$ is the
spectral function of the spin-down state
\footnote{Theoretical predictions for RF spectroscopy of single spin-orbit coupled bound fermion pair and
of noninteracting spin-orbit coupled Fermi gas have been reported in Ref.~\cite{Liu:2012}. Recently, the RF spectroscopy
of equal Rashba and Dresselhaus spin-orbit coupled Fermi gases has been studied experimentally and theoretically in Ref.~\cite{Hu:2013}. The theory part is based on a formalism
similar with what we used here.}.
Note that
\begin{eqnarray}
\int_{-\infty}^\infty d\o \G(\o)=1,
\end{eqnarray}
because
\begin{eqnarray}
n_\da=\frac{1}{V}\int_{-\infty}^\infty d\o \sum_\bk \ca_{\da\da}(\bk,\o)n_F(\o).
\end{eqnarray}

In \fig{figrf}, we present the integrated RF spectra for three different SOCs at resonance and at $T_c$, $T^*$, and $2T^*$.
The RF spectra are calculated in an idealized manner (see Appendix \ref{rfform}), \ie, we neglect the
final state interaction and width of the uncondensed pairs. Taking into
account the effects of finite width may change the shape of the RF spectra,
however, most of the qualitative features shall retain.

It is seen from \fig{figrf} that the RF spectra consist of two continuum branches, one positive and another
negative. The positive branches correspond to the ``binding" fermion pairs contribution, while
the negative branches can be regarded as the response of the thermal excited quasi-particles
with a pseudogap $\D_\pg$ reflected in the positions of the negative branch peaks. With increasing temperature, more quasi-particles are excited,
leading to a much more pronounced response appearing in the negative
branches. 
Such a temperature-sensitive feature of the RF spectroscopy may provide a useful way to experimentally measure the critical
temperature and to detect the existence of pseudogap in the normal phase.

\section {Summary}\label{discu}
We have theoretically investigated thermal effects on the BCS-BEC crossover of spin-orbit coupled
Fermi gases. For this purpose, we have employed a $T$-matrix formalism based on a $G_0G$ approximation for the
pair susceptibility, which was thoroughly used in the previous studies of Fermi gases without SOC. This
formalism extends the standard BCS theory by appropriately decomposing the excitation gap to a condensation part and
a pseudogap part that characterize the pairing fluctuations.

Comparing to the BCS theory, our $G_0G$ formalism predicts lower and more reliable critical temperature $T_c$ for the superfluid-normal
phase transition. The results for $T_c$ have been presented in \fig{tca} to \fig{tcl}. At various molecular BEC limits,
our predictions correctly recover the BEC temperature $T_\bec$ of free Bose gases. The pseudogap persists not only in the superfluid phase ($T<T_c$)
but also in a window of the normal phase ($T_c<T<T^*$) where it represents the existence of non-condensed, preformed pairs which
dissociate above $T^*$. We have studied how the SOC influences the emergence of the pseudogap, as shown
in \fig{dtl0}-\fig{dtuni}. It is seen that strong S- or EO- type SOC can significantly enlarge the pseudogap window in the normal phase.
Thus, spin-orbit coupled Fermi gases provide a new platform to study the pseudogap physics. Experimentally, such pseudogap might be revealed by
RF spectroscopy measurements. We have presented our qualitative predictions on the RF spectra in \fig{figrf}, which may be easily tested in future experiments.

{\bf Acknowledgments:}
LH and XGH are supported by the Helmholtz International Center for FAIR within the
framework of the LOEWE program (Landesoffensive zur Entwicklung
Wissenschaftlich- \"Okonomischer Exzellenz) launched by the State of Hessen. XGH is also
supported by Indiana University Grant No. 22-308-47 and the US DOE Grant No. DE-FG02-87ER40365.
XJL and HH are supported by the ARC Discovery Projects DP0984637 and DP0984522.

\appendix
\section {Expressions for the RF spectroscopy}\label{rfform}
In this appendix, we list some expressions for the idealized RF spectroscopy for Fermi gases with and without SOC. Some of them
are used in Sec.\ref{rf}.
By ``idealized", we mean that these expressions neglect the effects due to final state interactions and due to the finite
lifetime effects of the uncondensed pairs.
\begin{widetext}
(I)Without SOC. The spectral function of spin-down fermion is given by
\begin{eqnarray}
\ca^0_{\da\da}(\bk,\o)= u_\bk^2\d(\o-E_\bk)+v_\bk^2\d(\o+E_\bk).
\end{eqnarray}
The RF spectrum:
\begin{eqnarray}
\G_0(\o)=\frac{m^{3/2}}{4\p^2}\frac{\D^2}{\o^2}\sqrt{\frac{\o^2-\D^2+2\m\o}{\o}}n_F\lb-\frac{\o^2+\D^2}{2\o}\rb
\H\lb\frac{\o^2-\D^2+2\m\o}{\o}\rb.\non
\end{eqnarray}
\\ (II) S SOC. The spectral function spin-down fermion is given by
\begin{eqnarray}
\ca^{\rm S}_{\da\da}(\bk,\o)&=& \frac{1}{2}\sum_\a\lb1-\a\frac{k_z}{|\bk|}\rb
\ls (u^\a_\bk)^2\d(\o-E^\a_\bk)+(v^\a_\bk)^2\d(\o+E^\a_\bk)\rs.
\end{eqnarray}
The integrated RF spectrum:
\begin{eqnarray}
\G_{\rm S}(\o)&=&\frac{1}{4\p^2n_\da}\sum_\a\int_0^\infty d|\bk| |\bk|^2\ls (u^\a_\bk)^2\d(\x_\bk-\o-E^\a_\bk)+(v^\a_\bk)^2\d(\x_\bk-\o+E^\a_\bk)\rs n_F(\x_\bk-\o).
\end{eqnarray}
\\ (III) EO SOC. The spectral function of spin-down fermion is given by
\begin{eqnarray}
\ca^{\rm EO}_{\da\da}(\bk,\o)&=& \frac{1}{2}\sum_\a
\ls (u^\a_\bk)^2\d(\o-E^\a_\bk)+(v^\a_\bk)^2\d(\o+E^\a_\bk)\rs.
\end{eqnarray}
The integrated RF spectrum:
\begin{eqnarray}
\G_{\rm EO}(\o)&=&\frac{\sqrt{2m}}{16\p^2n_\da}\sum_\a\int_0^\infty dk_\perp k_\perp
\frac{\D^2}{(\o+\a\l k_\perp)^2}
n_F\ls-
\frac{\D^2+(\o+\a\l k_\perp)^2}{2(\o+\a\l k_\perp)}\rs
\ls\frac{\o^2-\D^2-\l^2k_\perp^2}
{2(\o+\a\l k_\perp)}
-\frac{k_\perp^2}{2m}+\m\rs^{-1/2}\non&&\times\H\ls\frac{\o^2-\D^2-\l^2k_\perp^2}
{2(\o+\a\l k_\perp)}
-\frac{k_\perp^2}{2m}+\m\rs.
\end{eqnarray}
\\ (IV) EP SOC. The spectral function of spin-down fermion is given by:
\begin{eqnarray}
\ca^{\rm EP}_{\da\da}(\bk,\o)&=& \frac{1}{2}\sum_\a\lb1-\a\frac{k_z}{|\bk|}\rb
\ls (u^\a_\bk)^2\d(\o-E^\a_\bk)+(v^\a_\bk)^2\d(\o+E^\a_\bk)\rs.
\end{eqnarray}
The integrated RF spectrum:
\begin{eqnarray}
\G_{\rm EP}(\o)&=&\frac{m^2}{8\p^2n_\da}\sum_\a\int_0^\infty dk_z \frac{\D^2}{(\o+\a\l k_z)^2}
n_F\ls-
\frac{\D^2+(\o+\a\l k_z)^2}{2(\o+\a\l k_z)}\rs\H\ls\frac{\o^2-\D^2-\l^2k_z^2}{2(\o+\a\l k_z)}
-\frac{k_z^2}{2m}+\m\rs.
\end{eqnarray}
\end{widetext}

\end{document}